\documentclass[floatfix,aps,prd,eqsecnum,showpacs,amsmath,amssymb,nofootinbib,preprintnumbers]{revtex4}
\pdfoutput=1
\usepackage{graphicx}
\usepackage{color}

\usepackage{hyperref}
\hypersetup{colorlinks=true,linkcolor=blue,citecolor=blue,pdfstartview=XYZ}

\usepackage{tikz}

\def\laq{~\raise 0.4ex\hbox{$<$}\kern -0.8em\lower 0.62ex\hbox{$\sim$}~}
\def\gaq{~\raise 0.4ex\hbox{$>$}\kern -0.7em\lower 0.62ex\hbox{$\sim$}~}

\def\beq{\begin{equation}}
\def\eeq{\end{equation}}
\def\bea{\begin{eqnarray}}
\def\eea{\end{eqnarray}}

\def\re {\rangle}

\def \pa {\partial}

\def \ti {\widetilde}

\def \La {\Lambda}
\def \Da {\Delta}
\def \da {\delta}

\def \ga {\gamma}
\def \sg {\sigma}
\def \Sg {\Sigma}
\def \da {\delta}

\def \r {\rho}

\def \Om {\Omega}

    \def\be{\begin{equation}}
    \def\ee{\end{equation}}
    \def\ba{\begin{eqnarray}}
    \def\ea{\end{eqnarray}}

    \def\e{\mbox{e}}

\newcommand{\eq}{\begin{equation}}
\newcommand{\eqx}{\end{equation}}
\newcommand{\eqn}{\begin{eqnarray}}
\newcommand{\eqnx}{\end{eqnarray}}
\usepackage{xspace}

\newcommand{\Ups}{\Upsilon}

\newcommand{\etal}{\emph{et al.}\xspace}

\newcommand{\rref}[1]{(\ref{#1})}

\newcommand{\lla}{\left\langle}
\newcommand{\rra}{\right\rangle}

\newcommand{\Acal}{\mathcal A}

\newcommand{\Ccal}{\mathcal C}
\newcommand{\Ocal}{\mathcal O}
\newcommand{\Tcal}{\mathcal T}
\newcommand{\Hcal}{\mathcal H}
\newcommand{\Ical}{\mathcal I}
\newcommand{\Pcal}{\mathcal P}

\def \L{\rm L}
\def \NL{\rm NL}

\usepackage{cancel}

\usepackage{pifont}

\begin{document}

\preprint{BA-TH/666-12}
\preprint{CERN-PH-TH/2012-362}
\preprint{LPTENS-13/01}
\preprint{DESY 13-011}

\title{ Average and dispersion of the luminosity-redshift relation \\ in the  concordance model} 

\author{I. Ben-Dayan$^{1}$, M. Gasperini$^{2,3}$, G. Marozzi$^{4,5}$, F. Nugier$^{6}$ and G. Veneziano$^{4,7,8}$}

\affiliation{$^1$Deutches Elektronen-Synchrotron DESY, Theory Group, D-22603 Hamburg, Germany\\
$^{2}$Dipartimento di Fisica, Universit\`{a} di Bari, Via G. Amendola
173, 70126 Bari, Italy\\
$^{3}$Istituto Nazionale di Fisica Nucleare, Sezione di Bari, Bari, Italy\\
$^{4}$ Coll\`ege de France, 11 Place M. Berthelot, 75005 Paris, 
 France\\
$^{5}$ Universit\'e de Gen\`eve, D\'epartement de Physique Th\'eorique and CAP,
24 quai Ernest-Ansermet, CH-1211 Gen\`eve 4, Switzerland \\
$^{6}$ Laboratoire de Physique Th\'eorique de l'\'Ecole Normale Sup\'erieure, CNRS UMR 8549, 24 Rue Lhomond, 75005 Paris, France\\
$^{7}$CERN, Theory Unit, Physics Department, CH-1211 Geneva 23, Switzerland\\
$^{8}$ Center for Cosmology and Particle Physics, 
Department of Physics, New York University \\
4 Washington Place, New York, NY 10003,  USA }

\begin{abstract}
Starting from the luminosity-redshift relation recently  given up to second order in the Poisson gauge, we calculate the effects of the realistic stochastic background of perturbations of the so-called concordance model on the combined light-cone and ensemble average of various functions of the luminosity distance, and on their variance, as functions of redshift. We apply a gauge-invariant light-cone  averaging prescription which is free from infrared and ultraviolet divergences, making our results robust with respect to changes of the corresponding cutoffs. Our main conclusions, in part already anticipated in a recent letter for the case of a perturbation spectrum computed in the linear regime, are that  such inhomogeneities not only cannot avoid the need for dark energy, but also cannot prevent, in principle, the determination of its parameters down to an accuracy of order $10^{-3}-10^{-5}$, depending on the averaged observable and on the regime considered for the power spectrum. However, taking into account the appropriate corrections arising in the non-linear regime, we predict an irreducible scatter of the data approaching the $10\%$  level which, for limited statistics,  will necessarily limit  the attainable precision. The predicted dispersion appears to be in good agreement with current observational estimates of the distance-modulus variance due to Doppler and  lensing effects (at low and high redshifts, respectively), and represents a challenge for future precision measurements.

\end{abstract}

\vspace {1cm}~


\maketitle

\section {Introduction}
\label{Sec1}
\setcounter{equation}{0}

In a recent letter \cite{BGMNV2} we have presented the main ideas  and  most significant results of a preliminary study of the effects of a stochastic background of inhomogeneities on the determination of the dark-energy parameters in the context of modern precision cosmology. The main conclusions of that analysis, based on the use of a perturbation spectrum valid in the linear regime, were as follows. On the one hand, such kind of perturbations cannot simulate a substantial fraction of dark energy: their contribution to the averaged flux-redshift relation is both too small (at large values of the redshift $z$) and has the wrong $z$-dependence.  On the other hand, stochastic fluctuations add a new and relatively important dispersion with respect to the prediction of the homogeneous and isotropic Friedmann-Lema\^itre-Robertson-Walker (FLRW) cosmology. This dispersion is independent of the experimental apparatus, of the observational procedure, of the intrinsic fluctuations in absolute luminosity, and may prevent a determination of the dark-energy parameter $\Omega_{\Lambda}(z)$ down to the percent level -- at least if we are using the luminosity-redshift relation alone. Another important conclusion was that (light-cone averages of) different functions of the same observable get biased in different ways, with the energy flux sticking out as the observable which gets minimally affected by inhomogeneities, irrespectively of the redshift binning utilized\footnote{Also redshift binning reduces biases and selects the flux as the preferred variable \cite{Wang}.}. We should recall here that other possible sources of uncertainty, bias and scatter in the Hubble diagram have been studied in many previous papers (see e.g. \cite{3}-\cite{7}).

The method we have followed, in order to arrive at the above-mentioned conclusions, consists of two different steps. In the first step we start from an exact expression for the luminosity-redshift relation in the special ``geodesic light-cone"  gauge introduced in \cite{GMNV}. We then transform this expression, up to second order in perturbation theory, to another gauge in which perturbations are known up to that order, the so-called Poisson gauge (PG) (see e.g. \cite{PG}).
The second step consists of performing the relevant light-cone and ensemble averages, and in inserting a realistic power spectrum of stochastic perturbations. The light-cone average procedure  appropriate to this context was formulated and discussed in \cite{GMNV}, by extending to null hypersurfaces the gauge-invariant procedure for space-like domains previously defined in \cite{GMV1,GMV2} and also applied in \cite{Marozzi}. 

Details on the first stage of this two-step process have been presented in a recent paper  \cite{BMNV}, while in this work we provide a detailed implementation of the second step. The computation method is basically the same as the one already used  in \cite{BGMNV1}, but it will involve the full second-order results obtained in \cite{BMNV}. We will consider in detail both a Cold Dark Matter (CDM) model and a   $\Lambda$CDM 
one (the so-called concordance model). We will reproduce, in particular, the results reported in \cite{BGMNV2} based on the use of the power spectrum of \cite{Eisenstein:1997ik}, valid in the linear perturbative regime. However, we will also extend our treatment by adding the effects of baryons and by considering two parametrizations of the HaloFit model \cite{Smith:2002dz,Takahashi}, describing the density power spectrum in the non-linear regime.

The paper is organized as follows. In Sect. \ref{Sec2} we recall, for the sake of completeness, the results of \cite{BMNV} for the contribution of scalar perturbations to the light-cone average of the flux-redshift relation, to second order,  in the Poisson gauge. We also reorganize the many different contributions in a convenient form for the actual estimates to be carried out. In Sect. \ref{Sec3} we present a few important aspects and consequences of the process of combining  light-cone and ensemble averages, considering in particular the luminosity distance and its phenomenologically  most relevant functions. We also introduce a convenient spectral parametrization of the inhomogeneous averaged terms. In Sect. \ref{Sec4} we discuss some relevant results about the dynamical evolution of scalar perturbations, up to second order, required for the computation of their averaged contribution. In Sect. \ref{Sec5}, as a warm-up exercise, we apply our methods to a simple perturbed CDM model, where most calculations -- except for the explicit mode integration over the given power spectrum -- can be done analytically. In the relevant range of $z$, the full leading result can be written in terms of an explicit (and simple) function of $z$ times a particular moment of the  spectrum, and this shows that such a model badly fails  in explaining the data, both in magnitude (in particular, at large redshift) and in $z$-dependence. 

We then turn to the case of a perturbed $\Lambda$CDM model, where calculations are more involved. They are simplified by restricting our attention to the so-called ``enhanced terms" (i.e the dominant ones in the relevant range of  $z$), already  identified  in the CDM case.  In order to discuss a realistic perturbation background,  we will also consider a power spectrum which includes the contribution of baryonic matter and takes into account the ``Silk-damping" effect. In  Sect. \ref{Sec6} we restrict our computations to the linear power-spectrum proposed in \cite{Eisenstein:1997ik}, and we evaluate the impact of  stochastic inhomogeneities not only on the averaged flux-redshift relation, but also on other functions of the luminosity distance (in particular, on the distance modulus used in the analyses of the Supernovae data). We also  discuss  the dispersion induced by the presence of the perturbations. In Sect. \ref{Sec7} we take into account the effects of the non-linear regime by using the HaloFit parametrizations of \cite{Smith:2002dz,Takahashi}. We also compute the variance/dispersion expected in the distance modulus and attempt a first comparison with available data and phenomenological fits, in particular for what concerns lensing at large redshifts. 
 Finally,  in Sect. \ref{Sec8} we summarize our result and draw some conclusions. In addition, in Appendix A we discuss why vector and tensor perturbations, although interesting on their own, do not contribute to the averages discussed in this paper. In Appendix B we report the explicit results for various spectral coefficients used in the analysis of the CDM model.

%%%%%%%%%%%%%%%%%%%%%%%%%%%%%%%%%%%%%%%%%%%%%%%%%%%%%%%%%%%%%%%%%%%%%%

\section{Averaging the luminosity flux  at second order}
\label{Sec2}
\setcounter{equation}{0}

In this section we recall previous results on the light-cone average of the luminosity flux, $\langle \Phi \rangle \sim \langle d_L^{-2}\rangle$ (where $d_L$ is the luminosity distance), first computed in a generally inhomogeneous metric background, and then specialized to the case of a spatially flat FLRW metric perturbed, to second order, by the presence of small fluctuations of scalar, vector and tensor type.

\subsection{Exact expression for $\langle d_L^{-2}\rangle$ in the geodesic light-cone gauge}
\label{Sec2A}

When describing the  propagation of light emitted by sources lying on the past light-cone of a given observer, it is convenient to identify the null hypersurfaces along which the photons reach the observer with those on which a null coordinate takes constant values. For this reason  we have  introduced in \cite{GMNV}  an adapted system of coordinates -- defining what  we have called the  ``geodesic light-cone'' (GLC) gauge -- in which several quantities greatly simplify,   while keeping all the required degrees of freedom for applications to general geometries.  The coordinates $x^\mu= (\tau, w, \ti{\theta}^a)$ (with $a=1,2$, $\ti{\theta}^1 = \ti{\theta}$, $\ti{\theta}^2 = \ti{\phi}$), specifying the metric in the GLC gauge, correspond to a complete gauge fixing of the so-called observational coordinates,  defined e.g. in \cite{Maartens1,Maartens3}   (see also \cite{Clarkson}). 

The GLC metric depends indeed on six arbitrary functions (a function $\Ups$, a two-dimensional ``vector" $U^a$ and a symmetric matrix $\gamma_{ab}$),
and its line-element takes the form\footnote{We have put tildas on the GLC gauge $\theta^a$ coordinates in order to be consistent with our previous notations in \cite{BGMNV2,BMNV, BGMNV1}.}: 
\bea
\label{LCmetric}
ds_{GLC}^2 = \Ups^2 dw^2-2\Ups dw d\tau+\gamma_{ab}(d\ti{\theta}^a-U^a dw)(d\ti\theta^b-U^b dw) ~~.
 \eea
In matrix form,  the metric and its inverse are then given by:
\eq
\label{GLCmetric}
g^{GLC}_{\mu\nu} =
\left(
\begin{array}{ccc}
0 & - \Ups &  \vec{0} \\
-\Ups & \Ups^2 + U^2 & -U_b \\
\vec0^{\,T}  &-U_a^T  & \gamma_{ab} \\
\end{array}
\right)
~~~~~,~~~~~
g_{GLC}^{\mu\nu} =
\left(
\begin{array}{ccc}
-1 & -\Ups^{-1} & -U^b/\Ups \\
-\Ups^{-1} & 0 & \vec{0} \\
-(U^a)^T/\Ups & \vec{0}^{\, T} & \gamma^{ab}
\end{array}
\right) ~~~,
\eqx
where $\vec 0=(0,0)$, $U_b= (U_1, U_2)$, while the $2 \times 2$ matrices 
$\gamma_{ab}$ and  $\gamma^{ab}$ lower and  raise  the two-dimensional indices. Clearly $w$ is a null coordinate (i.e.  $\pa_\mu w \pa^\mu w=0$), and a past light-cone hypersurface is specified by the condition $w=$ const. We also note that $u_\mu \sim \partial_{\mu} \tau$ defines a geodesic flow, i.e. that
$\left( \pa^\nu \tau\right) \nabla_\nu \left( \pa_\mu \tau\right) = 0$
(as a consequence of   $g^{\tau \tau} = -1$). Such a 4-velocity defines  geodetic observers corresponding to the static ones in the synchronous gauge \cite{BGMNV1}. Let us also remark that, in GLC coordinates, the null geodesics connecting sources and observer are characterized by the simple tangent vector $k^{\mu} = g^{\mu \nu} \partial_{\nu} w =  g^{\mu w} = - \delta^{\mu}_{\tau} \Ups^{-1}$, meaning that photons  travel at constant $w$ and $\ti \theta^a$. This makes the calculation of  the redshift and of the area distance particularly easy in this gauge.

In fact, let us denote by the subscripts $o$ and $s$, respectively, a quantity evaluated at the observer and source space-time position, and consider a light ray emitted by a  geodetic source (with four-velocity $u_\mu=- \pa_\mu \tau$) lying at the intersection between the past light-cone of a given geodetic observer (defined by the equation $w=w_o$) and the spatial hypersurface $\tau= \tau_s$, where $\tau_s$ for the moment is a constant parameter. The light ray will be received by our static geodetic observer at $\tau=\tau_o>\tau_s$. The redshift $z_s$ associated with this  light ray is then given by \cite{GMNV}:
\be
\label{redshift}
(1+z_s) = \frac{(k^{\mu} u_{\mu})_s }{(k^{\mu} u_{\mu})_o}  = \frac{(\partial^{\mu}w \pa_\mu \tau)_s }{(\partial^{\mu}w \pa_\mu \tau)_o}  = {\Ups(w_o, \tau_o, \ti \theta^a)\over \Ups(w_o, \tau_s, \ti \theta^a)} ~~.
\ee

We shall be interested in averaging the luminosity at fixed redshift, hence on
 the two-dimensional surface $\Sigma(w_o, z_s)$ (topologically a sphere) which lies on our past light-cone ($w=w_o$)  and is associated with a fixed redshift ($z = z_s$). In terms of the $\tau$ coordinate such a surface corresponds to the equation $\tau = \tau_s(w_o, z_s, \tilde{\theta}^a)$ enforcing Eq. (\ref{redshift}). Hereafter $\tau_s$ will denote this (in general angle-dependent) quantity. 
 
Also the area distance $d_A$, related to the luminosity distance $d_L$ of a source at redshift $z_s$ by one of  Etherington's  relations \cite{Et1933}:
\be
 d_A = (1+z_s)^{-2} d_L ~~,
 \label{dLdA}
\ee
 takes a particularly simple form in the GLC gauge. A direct derivation \cite{GMNV} starts from its general definition \cite{Sachs}: 
\be
\label{dAGLC}
d_A^2 = \frac{dS_s}{d\Omega_o}  ~~,
\ee
where  $d\Omega_o $ is the infinitesimal solid angle subtended by the source at the observer position, and $dS_s$ is the area element  on the surface orthogonal to both the photon momentum and to the source 4-velocity at the source's position. It is easy to check that the surface $dS_s$  is characterized by having constant $w$ and $\tau$, and that the induced 2-metric on it is nothing but $\gamma_{ab}$  \cite{BMNV}. Therefore\footnote{This result has been checked meanwhile \cite{FGMV} by explicitly constructing the relevant Jacobi map in the GLC gauge and by using its known relation to the area distance \cite{Ehlers}. This method confirms the results discussed below (and in our previous papers), modulo a  Lorentz  transformation of $d\Omega_o$ connected to the peculiar velocity of the observer as measured in the longitudinal gauge. A similar correction is also needed, of course, in order to take into account the peculiar motion of our galaxy hence, strictly speaking, our unintegrated results only hold without these peculiar-velocity-related effects. In practice, the correction exactly vanishes for the averaged flux (Eq. (\ref{Phitheory}) below) and is numerically negligible for the  other averages discussed in our papers.}:
 \be
\label{dAGLC1}
d_A^2 = \frac{d^2 \tilde{\theta} \sqrt{\gamma}}{d^2 \tilde{\theta} \sin \tilde{\theta}} = \frac{\sqrt{\gamma}}{\sin \tilde{\theta}} ~~,
\ee
where we have used the fact that photons travel at constant $ \tilde{\theta}^a$. Our averaging surface $\Sigma(w_o, z_s)$, being one of constant $z$, differs from the one of constant $w$ and $\tau$ but -- amusingly -- the same formula holds, locally, for the area element on it, so that Eq. (\ref{dAGLC1}) can also be used for our light-cone averages  \cite{BMNV}.

The above result singles out the received luminosity flux, $\Phi \sim d_L^{-2} = (1+z_s)^{-4} d_A^{-2}$, as an  important -- and extremely simple --  observable to average over the 2-sphere $\Sigma(w_o, z_s)$ embedded in the light-cone. In fact (see \cite{BMNV} for more details):
\beq 
\label{Phitheory}
\langle d_L^{-2} \rangle(w_o,z_s)=(1+z_s)^{-4}\frac{\int dS \frac{d\Omega_o}{dS}}{\int dS}=(1+z_s)^{-4}\frac{\int d\Omega_o}{\int dS}=(1+z_s)^{-4}\frac{4\pi}{\Acal(w_o, z_s) } ~~,
\eeq
where
\beq
\Acal(w_o, z_s) =  \int _{\Sigma(w_o,z_s)} d^2 \tilde{\theta}^a  \sqrt{\ga} ~~
\eeq
is the proper area of $\Sigma(w_o,z_s)$ computed with the metric $\gamma_{ab}$, and expressed in terms of internal coordinates $(w_o,z_s)$ parametrizing the deformed 2-sphere $\Sigma(w_o,z_s)$.

Eq. (\ref{Phitheory}) holds non-perturbatively for any space-time geometry, and is the starting point for the computation of the average flux summarized in \cite{BGMNV2} and presented in details here. By using the notations introduced in \cite{BGMNV2} we can write, in particular, 
\be
\label{dLminus2}
\langle d_L^{-2} \rangle(w_o, z_s) = (1+z_s)^{-4} \left[ \int \frac{d^2 \tilde{\theta}^a}{4 \pi} \sqrt{\gamma}(w_o, \tau_s(w_o, z_s,\tilde{\theta}^a), \tilde{\theta}^a)  \right]^{-1} \equiv \left(d_L^{FLRW}\right)^{-2} I_\phi^{-1} (w_o, z_s)~~,
\ee
where we have defined 
\be
\label{Integral}
I_\phi(w_o, z_s) \equiv \frac{{\cal A}(w_o, z_s)}{4 \pi \left[a(\eta_s^{(0)}) \Delta \eta \right]^2},
\ee
and where $d_L^{FLRW}= (1+z_s)^2 a(\eta_s^{(0)}) \Delta \eta$ is the luminosity distance for the unperturbed FLRW geometry, with scale factor $a(\eta)$. Here $\eta$ is the conformal time coordinate, $\Da \eta = \eta_o-\eta_s^{(0)}$, and we have denoted with $\eta_s^{(0)}$ the background solution of the equation for the source's conformal time  $\eta_s= \eta_s(z_s, \ti \theta^a)$ (see \cite{BMNV,BGMNV1}). Note that, according to the above equation,  the interpretation of $I_\phi(w_o, z_s)$ is straightforward: it is simply the ratio of the area of the 2-sphere at redshift $z_s$ on the past light-cone (deformed by inhomogeneities), over the area of the corresponding homogeneous 2-sphere.

%%%%%%%%%%%%%%%%%%%%%%%%%%%%%%%%%%%%%%%%%%%%%%%%%%%%%%%%%%%%%%%%%%%%%%%%%%%%%%%%%%%%%%%%%%%%%%%%%%%%

\subsection{Second-order expression for $\langle d_L^{-2}\rangle$ in the Poisson gauge}
\label{Sec2B}

Let us consider a  space-time geometry that can be approximated by a spatially flat FLRW metric distorted by the presence of scalar, vector and tensor perturbations. In the so-called Poisson gauge (PG) \cite{PG} (a generalization of the standard Newtonian gauge beyond first order),  the corresponding metric (in Cartesian coordinates) takes the form:
\be
ds_{PG}^2 = a^2(\eta) \left\{ -(1+ 2 \Phi) d\eta^2 + 2 \omega_i d\eta d x^i + \left[(1- 2 \Psi)\delta_{ij} + h_{ij}   \right] dx^i dx^j \right\} \,.
\label{PGmetricstandard}
\ee
Here $\Phi$ and $\Psi$ are scalar perturbations, $\omega_i$ is a transverse vector perturbation ($\partial^i \omega_i=0$) and $h_{ij}$ is a transverse and traceless tensor perturbation ($\partial^i h_{ij}=0=h^i_i$). This metric depends on six arbitrary functions, hence it is completely gauge fixed. By including first-order and second-order contributions,  the (generalized) Bardeen potentials $\Phi$ and $\Psi$ can be defined as follows:
\be
\label{PsiPhiUpTo2}
\Phi \equiv \psi + \frac{1}{2} \phi^{(2)} ~~,~~ \Psi \equiv \psi + \frac{1}{2} \psi^{(2)} \,,
\ee
where we have assumed the absence of anisotropic stress in order to set $\Psi = \Phi= \psi$ at first order. 

It is important to stress, at this point, that for the purpose of this paper we can safely restrict our subsequent discussion to the case of pure scalar perturbations.  In fact, it is true that at second order different perturbations get mixed: vector and tensor perturbations are automatically generated from scalar perturbations  (see e.g. \cite{Matarrese:1997ay,BMR}), while second-order scalar perturbations are generated from first-order vector and tensor perturbations. 
However, a single vector or tensor perturbation does not contribute to our angular averages on $\Sigma(w_o,z_s)$  (see Appendix A for further details). Furthermore,  we will treat $\omega_i$ and $h_{ij} $ as second order quantities. In other words, we assume the first-order perturbed metric to be dominated by scalar contributions (which is indeed the case if perturbations are generated by a phase of standard slow-roll inflation, see e.g. \cite{DurCos,MG}). As a result, we shall also neglect the contributions induced, at second order, 
by first-order vector and tensor perturbations. 
 
We have already established in \cite{BMNV}  a connection  between the second-order perturbative expression of the luminosity distance $d_L$ and of  the integrand of $I_\phi$ (controlling $\langle d_L^{-2}\rangle$), both written in terms of the PG perturbations and of the observer's angles $\tilde{\theta}^a$ (remember that photons reach the observer traveling at constant $\tilde{\theta}^a$). In particular, by defining\footnote{Note the simplified notation with respect to
\cite{BMNV}. We have also omitted the indication that $w = w_o$.} 
\beq
\frac{{d}_L(z_s, \tilde{\theta}^a)}{(1+z_s)a_o \Delta \eta}
= {{d}_L(z_s, \tilde{\theta}^a)\over d_L^{FLRW}(z_s)} = 1 + \delta_S^{(1)}(z_s, \tilde{\theta}^a) + \delta_S^{(2)}(z_s, \tilde{\theta}^a) ~~, 
\label{215}
\eeq
and
\be
\label{Iphi}
I_\phi(z_s)=\int \frac{d^2 \tilde{\theta}^a}{4 \pi} \sin \tilde{\theta}
\left(1+{\cal I}_1+{\cal I}_{1,1}+{\cal I}_2\right) ~~,
\ee
we have found that \cite{BMNV}\footnote{As already mentioned in a previous footnote, the terms appearing on the r.h.s. of  Eq.(\ref{215}) should be corrected, in principle, for the change in $d\Omega_o$ stemming from the peculiar velocity of the observer. However, such a correction has no effect on the integral $I_{\phi}$ itself, since it can be compensated by a Lorentz transformation of the angular variables.}
\bea
\label{Ideltarel}
{\cal I}_1 = 2 {\delta}_S^{(1)} + (\rm{t.~d.})^{(1)}, ~~~~~~~~~~~
{\cal I}_{1,1}+{\cal I}_2 = 2  {\delta}_S^{(2)}  + ( {\delta}_S^{(1)})^2 +  (\rm{t.~d.})^{(2)}\, ,
\eea
where the $ (\rm{t.~d.})^{(1,2)}$ denote total derivatives terms w.r.t. the $\tilde{\theta}^a$ angles,  giving vanishing contributions either by periodicity in $\ti{\phi}$ or by the vanishing of the integrand at $\ti{\theta} = 0, \pi$. We only recall here, for later use, the first-order total derivative:
\be
({\rm t.~d.})^{(1)}= 2\, J_2^{(1)}~~, ~~~~~~~~~J_2^{(1)} = \frac{1}{\Delta\eta} \int_{\eta_s^{(0)}}^{\eta_o} d \eta \,\frac {\eta - \eta_s^{(0)}}{\eta_o - \eta} \Delta_2 \psi(\eta, \eta_o-\eta, \ti \theta^a),
\label{216}
\ee
where  $\Delta_2= \pa^2_{\ti\theta} + \cot \ti\theta \pa_{\ti\theta} + \sin^{-2} \ti\theta \pa^2_{\ti\phi}$. It is also convenient to rewrite the PG metric using spherical coordinates (but still considering that photons travel at constant $\tilde{\theta}^a$), and define the following quantities \cite{BMNV}:
\bea
&&
P(\eta, r, \ti\theta^a) = \int_{\eta_{in}}^\eta d\eta' \frac{a(\eta')}{a(\eta)} \psi(\eta',r,\ti\theta^a) ~~,~~~~~~~~ Q(\eta_+, \eta_-, \ti\theta^a) = \int_{\eta_+}^{\eta_-} dx~ \hat{\psi}(\eta_+,x,\ti\theta^a) ~~,~~  \nonumber \\
&&
\Xi_s =  1 - \frac{1}{\Hcal_s \Delta \eta} ~~, ~~~~~~~~~~~~~~~~~~~~~~~~~~~~~~~~~~J  = \left([\partial_+ Q]_s - [\partial_+ Q]_o\right) - \left([\partial_r P]_s - [\partial_r P]_o\right)\,.
\eea
Here $\Hcal=d \ln a /d\eta$, and the lower limit $\eta_{in}$ represents an early enough time when the perturbation (or better the integrand) was negligible.  When ambiguities may occur the superscript  ${(0)}$ denotes the background solution of a given quantity (similarly, the superscripts  ${(1)}$,  ${(2)}$ will denote, respectively, the first- and second-order perturbed values of that quantity). In the above equations  we have also introduced the useful (zeroth-order) light-cone variables $\eta_\pm= \eta \pm r$, 
with corresponding partial derivatives:
\beq
\pa_\eta = \pa_+ + \pa_- ~~~,~~~~~ \pa_r = \pa_+ - \pa_- ~~~,~~~~~\pa_\pm= {\pa \over \pa \eta_\pm}={1\over 2} \left( \pa_\eta \pm \pa_r \right) ~~.
\eeq
We shall use hereafter a hat to denote a quantity expressed in terms of the $(\eta_+,\eta_-,\ti\theta^a)$ variables, so that, for instance, $\hat{\psi}(\eta_+,\eta_-,\ti\theta^a) \equiv \psi(\eta,r,\ti\theta^a)$. Finally, in order to understand the physical meaning of the above quantities, it may be helpful to recall that the radial gradient of $P$ is related to the Doppler effect (due to peculiar velocities of source and observer), while the gradient of $Q$ with respect to $\pa_+$ represents the Sachs-Wolfe and the integrated Sachs-Wolfe effect. The last term $J$ corresponds to a combination of the three mentioned effects (see \cite{BMNV}, Sect. 4, for a more detailed discussion).
 
The results obtained in \cite{BMNV} can then be reported in the following form:
\be
{\cal I}_{1} = \sum_{i = 1}^{3} \Tcal_i^{(1)} ~~~~~;~~~~~ {\cal I}_{1,1} = \sum_{i = 1}^{23} \Tcal_i^{(1,1)} ~~~~~;~~~~~ {\cal I}_{2} = \sum_{i = 1}^{7} \Tcal_i^{(2)}
~~~~~;
\ee
where ${\cal I}_1,~ {\cal I}_{1,1},~ {\cal I}_2 $ are, respectively, the first-order, quadratic first-order, and genuine second-order  contributions of our stochastic fluctuations, and where: 
\beq
\label{DefI1}
\Tcal_1^{(1)} = - 2 \psi (\eta_s^{(0)}, r_s^{(0)}, \ti\theta^a)
~;~~~~~~~~
\Tcal_2^{(1)} = 2 \Xi_s J
~~~;~~~~~~~
\Tcal_3^{(1)} = - \frac{2}{\Delta \eta} Q_s
~~;
\eeq
\bea
\label{DefI11part1}
& &
\Tcal_1^{(1,1)} = \Xi_s \left[ \psi_s^2 - \psi_o^2 \right]
~~;~~
\Tcal_2^{(1,1)} = \Xi_s \left( ([\partial_r P]_s)^2 - ([\partial_r P]_o)^2 \right)
~~;~~
\Tcal_3^{(1,1)} = - 2 \Xi_s \left(\psi_s + [\partial_+ Q]_s\right) [\partial_r P]_s
~~;~~ \nonumber \\
& &
\Tcal_4^{(1,1)} = \frac12 \Xi_s (\gamma_0^{ab})_s \left( 2 \partial_a P_s \partial_b P_s + \partial_a Q_s \partial_b Q_s - 4 \partial_a Q_s \partial_b P_s \right)
~~;~~
\Tcal_5^{(1,1)} = - \Xi_s \lim_{r\rightarrow 0} \left[\gamma_0^{ab} \partial_a P \partial_b P \right] 
~~;~~ \nonumber \\
& &
\Tcal_6^{(1,1)} = 2 \Xi_s Q_s \left( 2 \partial_r \psi_o + 2 \partial_\eta \psi_o - \partial_r \psi_s + 2 \int_{\eta_s^{(0)}}^{\eta_0}d \eta' \partial_r^2
\psi\left(\eta',\eta_o-\eta',\ti\theta^a\right) + [ \partial_r^2 P ]_s \right)
~~;~~ \nonumber \\
& &
\Tcal_7^{(1,1)} = 2 \Xi_s \frac{J}{{\mathcal H}_s} \left( [\partial_\eta \psi]_s + {\mathcal H}_s [\partial_r P]_s \right)
~~;~~
\Tcal_8^{(1,1)} = 2 \Xi_s \frac{J}{{\mathcal H}_s} [\partial_r^2 P]_s
~~;~~ \nonumber \\
& &
\Tcal_9^{(1,1)} = - \Xi_s \int_{\eta_{in}}^{\eta_s^{(0)}} d\eta' \frac{a(\eta')}{a(\eta_s^{(0)})} \partial_r \left[ - \psi^2 + (\partial_r P)^2 + \gamma_0^{ab} \partial_a P \partial_b P \right] (\eta', \Delta \eta, \tilde{\theta}^a)
~~;~~ \nonumber \\
& &
\Tcal_{10}^{(1,1)} = \Xi_s \int_{\eta_{in}}^{\eta_o} d\eta' \frac{a(\eta')}{a(\eta_o)}  \partial_r \left[ - \psi^2 + (\partial_r P)^2 + \gamma_0^{ab} \partial_a P \partial_b P \right] (\eta', 0, \tilde{\theta}^a)
~~;~~ \nonumber \\
& &
\Tcal_{11}^{(1,1)} = 2 \Xi_s \int_{\eta_s^{(0)+}}^{\eta_s^{(0)-}} dx~ \partial_+ \left[ \hat{\psi} ~ \partial_+ Q + \frac14 \hat{\gamma}_{0}^{ab} \partial_a Q \partial_b Q \right] (\eta_s^{(0)+}, x, \tilde{\theta}^a)
~~;~~
\nonumber
\eea
\bea
\label{DefI11part2}
& &
\Tcal_{12}^{(1,1)} = \left[ \Xi_s^2 - \frac{1}{{\mathcal H}_s \Delta \eta} \left( 1 - \frac{{\mathcal H}_s'}{{\mathcal H}_s^2} \right) \right] J^2
~~;~~
\Tcal_{13}^{(1,1)} = - 4 \psi_s J
~~;~~
\Tcal_{14}^{(1,1)} = 2 \Xi_s \left\{ \psi_o + [\partial_r P]_o - \frac{Q_s}{\Delta \eta} \right\} J
~~;~~ \nonumber \\
& &
\Tcal_{15}^{(1,1)} = - 2 (J - 2 \psi_s) \frac{Q_s}{\Delta \eta}
~~;~~
\Tcal_{16}^{(1,1)} = - 2 \left( \psi_s - \partial_+ Q_s \right) \frac{Q_s}{\Delta \eta}
~~;~~
\Tcal_{17}^{(1,1)} = \left(\frac{Q_s}{\Delta \eta}\right)^2
~~;~~ \nonumber \\
& &
\Tcal_{18}^{(1,1)} = \frac{1}{\Hcal_s} (\gamma_0^{ab})_s \partial_a Q_s \partial_b J
~~;~~
\Tcal_{19}^{(1,1)} = \frac12 (\gamma_0^{ab})_s \partial_a Q_s \partial_b Q_s
~~;~~
\Tcal_{20}^{(1,1)} = 2 \frac{J}{{\mathcal H}_s} (-[\partial_\eta \psi]_s + [\partial_r \psi]_s)
~~;~~ \nonumber \\
& &
\Tcal_{21}^{(1,1)} = 2 Q_s [\partial_r \psi]_s
~~;~~
\Tcal_{22}^{(1,1)} = - \frac{2}{\Delta\eta} \int_{\eta_s^{(0)+}}^{\eta_s^{(0)-}} dx~ \left[ \hat{\psi} ~ \partial_+ Q + \frac14 \hat{\gamma}_{0}^{ab} ~ \partial_a Q ~ \partial_b Q \right] (\eta_s^{(0)+},x,\tilde{\theta}^a)
~~;~~ \nonumber \\
& &
\Tcal_{23}^{(1,1)} =  \frac{1}{8\sin \tilde{\theta}} \frac{\partial}{\partial \tilde{\theta}} \left\{ \cos \tilde{\theta} ~ \left( \int_{\eta_s^{(0)+}}^{\eta_s^{(0)-}} dx~ [\hat{\gamma}_{0}^{1b} ~ \partial_b Q](\eta_s^{(0)+},x,\tilde{\theta}^a) \right)^2 \right\}~;
\eea
and
\bea
\label{DefI2}
& &
\Tcal_1^{(2)} = - \frac{1}{2} \Xi_s \left( \phi_s^{(2)} - \phi_o^{(2)} \right)
~~;~~
\Tcal_2^{(2)} = \frac{1}{2} \Xi_s \left( \psi_s^{(2)} - \psi_o^{(2)} \right)
~~;~~
\Tcal_3^{(2)} = - \Xi_s \int_{\eta_{in}}^{\eta_s^{(0)}} d\eta' \frac{a(\eta')}{a(\eta_s^{(0)})} [\partial_r \phi^{(2)}] (\eta', \Delta \eta, \tilde{\theta}^a)
~~;~~ \nonumber \\
& &
\Tcal_4^{(2)} = \Xi_s \int_{\eta_{in}}^{\eta_o} d\eta' \frac{a(\eta')}{a(\eta_o)} [\partial_r \phi^{(2)}](\eta', 0, \tilde{\theta}^a)
~~;~~
\Tcal_5^{(2)} = \frac12 \Xi_s \int_{\eta_s^{(0)+}}^{\eta_s^{(0)-}} dx~ \partial_+ \left[ \hat{\phi}^{(2)} + \hat{\psi}^{(2)} \right](\eta_s^{(0)+}, x, \tilde{\theta}^a)
~~;~~ \nonumber \\
& &
\Tcal_6^{(2)} = - \psi_s^{(2)}
~~;~~
\Tcal_7^{(2)} = - \frac{2}{\Delta\eta} \int_{\eta_s^{(0)+}}^{\eta_s^{(0)-}} dx~ \left[ \frac{\hat{\phi}^{(2)} + \hat{\psi}^{(2)}}{4} \right] (\eta_s^{(0)+},x,\tilde{\theta}^a)
~~.
\label{Tcal2}
\eea
In the above equations $\ga_0^{ab}= {\rm diag} (r^{-2}, r^{-2} \sin^{-2} \ti \theta)$,
and for $\Tcal_6^{(1,1)}$ we have used the following identity:
\be
 - [\partial_+^2 Q]_s + [\partial_+ \hat{\psi}]_s=2 \partial_r \psi_o + 2 \partial_\eta \psi_o - \partial_r \psi_s + 2 \int_{\eta_s^{(0)}}^{\eta_0}d \eta' \partial_r^2
\psi\left(\eta',\eta_o-\eta', \ti\theta^a\right) \, .
\ee

Let us  point out, finally, that the last term in Eq.(\ref{DefI11part2}) corresponds to a total derivative, and thus to a boundary contribution, that superficially looks non vanishing. We believe that this is  the result of a naive treatment of the angular coordinate transformation, which becomes singular near the poles of the 2-sphere. This contribution, indeed, has the same form as the irrelevant one coming from an overall $SO(3)$ rotation.

%%%%%%%%%%%%%%%%%%%%%%%%%%%
%%%%%%%%%%%%%%%%%%%%%%%%%%

\section{Combining the light-cone and ensemble average of $d_L$ and of its functions}
\label{Sec3}
\setcounter{equation}{0}

In the cosmological model we are considering, the deviations from the standard FLRW geometry are sourced by a stochastic background of primordial perturbations  satisfying $\overline \psi=0$, $\overline{\psi^2} \not=0$, where the bar denotes statistical (or {\em  ensemble}) average. Hence, non-trivial effects on the {\em  ensemble} average of $d_L$, or of a generic function of it, can only originate either from quadratic and higher-order perturbative corrections, or  from the spectrum of correlation functions such as $\overline{d_L(z, \ti\theta^a) d_L(z', \ti\theta^{\prime a})}$ (see \cite{Bonvin}).  In this paper, rather than considering the  {\em ensemble} average of $d_L$,  we shall deal with that of $\langle d_L \re$, where the angular brackets refer to the light-cone average defined in \cite{GMNV} and presented in Sect. \ref{Sec2} (see \cite{Li:2007ny,precision,CU} for previous attempts of combining {\em ensemble} and space-time averages in the case of {\em spacelike} hypersurfaces). 

As already stressed in \cite{BGMNV1}, given  the covariant (light-cone) average of a perturbed (inhomogeneous) observable $S$,  the average of a generic function of this observable  differs, in general, from the function of its average, i.e. $\overline{\langle F(S) \rangle} \not= F(\overline{\langle S\rangle})$ (as a consequence of the nonlinearity of the averaging process).
 Expanding the observable  to second order as $S=S_0+S_1+S_2$, 
 one finds \cite{BGMNV2}
\beq
\overline{\langle F(S) \rangle} = F(S_0)+ F'(S_0) \overline{\langle S_1+S_2  \rangle}+F''(S_0) \overline{\langle S_1^2/2 \rangle},
\label{3}
\eeq
where in general $\overline{\langle S_1\rangle}\neq 0$ as a consequence of the so-called ``induced backreaction" terms,  arising from the coupling between the inhomogeneity fluctuations of $S$ and those of the integration measure  (see \cite{BGMNV1}). The overall correction to $\overline{\langle F(S) \rangle}$ thus depends not only on the intrinsic inhomogeneity of the observable $S$, but also on the covariance properties of the adopted averaging procedure.  Eq. (\ref{3}) implies, in our case, that different functions of the luminosity distance (or of the flux) may be differently affected by the process of averaging out the inhomogeneities, and  may require different ``subtraction" procedures for an unbiased determination of the relevant observable quantities. 

Let us consider, in particular, the luminosity flux $\Phi \sim d_L^{-2}$
(not to be confused with the Bardeen potential!), computed in Sect. \ref{Sec2}. Performing the stochastic average of Eq. (\ref{dLminus2}) (and using Eq. (\ref{Iphi})) we obtain 
\beq
\overline{\langle d_L^{-2} \rangle}(z) = (d_L^{FLRW})^{-2}\overline{(I_\Phi(z))^{-1}} \equiv (d_L^{FLRW})^{-2} \left[1 + f_\Phi(z) \right],
\label{7}
\eeq
where:
\beq
f_\Phi(z) \equiv \overline{\lla \Ical_1 \rra^2} - \overline{\lla \Ical_{1,1} + \Ical_2 \rra} ~~.
\label{7a}
\eeq
We can now apply the general result (\ref{3}) to the flux variable, by setting $S= \Phi$  and considering two important functions of the flux:  $F(\Phi)= \Phi^{-1/2} \sim d_L$, and  $F(\Phi)= -2.5 \log_{10}\Phi + {\rm const} \sim \mu$ (the distance modulus). They will be considered in the following sections, together with the flux. 
For the luminosity distance we can introduce a fractional correction $f_d$, in analogy with Eq. (\ref{7}), such that:
\beq
\overline{\langle d_L \rangle}(z)=d_L^{FLRW} \left[1+f_d(z)\right] ~~.
\label{2}
\eeq
Then, by using the general expression (\ref{3}), we find:
\beq
f_d = -{1\over 2}f_\Phi+{3\over 8}\,\overline{\langle \left(\Phi_1/\Phi_0\right)^2 \rangle} ~~,
\label{13}
\eeq
where, in terms of the quantities defined in (\ref{215}), we have $\overline{\langle \left(\Phi_1/\Phi_0\right)^2 \rangle} = 4 \overline{\langle (\delta_S^{(1)})^2 \rangle}$, and where $f_\Phi$ is defined by Eq. (\ref{7a}).  
For the distance modulus we obtain, instead, 
\beq
\overline{\langle \mu \rangle}- \mu^{FLRW}= - 1.25 (\log_{10}e)
\Big[ 2f_\Phi-\overline{\langle \left(\Phi_1/\Phi_0\right)^2 \rangle}
\Big] ~~.
\label{14}
\eeq

We can also consider, for any given averaged variable $\langle S \rangle$, the associated dispersion $\sg_S$ controlling how broad is the distribution of a perturbed observable $S$ around its mean value $\overline{\langle S  \rangle}$. This dispersion is due to both the geometric fluctuations of the averaging surface and to the statistical {\em ensemble} fluctuations, and is defined, in general, by \cite{BGMNV1}:
 \begin{equation}
 \label{vargen}
\sigma_S \equiv \sqrt{\overline{\lla \left(S - \overline{\langle S \rangle} \right)^2 \rra}}  =\sqrt{ \overline{\lla S^2 \rra}  - \left(\overline{\lla S \rra}\right)^2} \,.
\end{equation}
The dispersion associated with the flux is thus given by:
\beq
\sigma_\Phi=\sqrt{ \overline{\langle \left(\Phi/\Phi_0\right)^2 \rangle}- \left( \overline{\langle \Phi/\Phi_0 \rangle}\right)^2}=  \sqrt{\overline{\langle \left(\Phi_1/\Phi_0\right)^2 \rangle}} ~~,
\label{16}
\eeq
while for the distance modulus we  find:
\beq
\sigma_\mu=\sqrt{ \overline{\langle \mu^2 \rangle}- \left( \overline{\langle \mu \rangle}\right)^2}=  2.5 (\log_{10} e) \sqrt{\overline{\langle \left(\Phi_1/\Phi_0\right)^2 \rangle}} ~~.
\label{15}
\eeq
The above results will be applied to the case of a realistic background of cosmological perturbations of inflationary origin in the following sections.

Let us conclude this section by introducing a convenient spectral parametrization to be used for the various terms contributing to the fractional corrections of our observables, and to the corresponding dispersions. 
We start by recalling that the simplest way to implement the {\em ensemble} average of a stochastic background of scalar perturbations $\psi$ is to consider its Fourier decomposition in the form:
\be
\psi(\eta, \vec{x}) = \frac{1}{(2 \pi)^{3/2}} \int d^3 k \, \e^{i\vec{k}\cdot \vec{x}} \psi_k(\eta) E(\vec{k}) \,,
\label{PsiFourier}
\ee
where -- assuming that the fluctuations are statistically homogeneous and isotropic -- $E$  is a unit random variable satisfying $E^*(\vec{k})=E(-\vec{k})$, as well as  the {\em ensemble}-average conditions $\overline{E(\vec{k})} = 0$ and $\overline{E(\vec{k}_1) E(\vec{k}_2)}=\delta(\vec{k}_1+\vec{k}_2)$.
As a simple  illustrative example one has:
\bea
\overline {\lla \psi_s \psi_s \rra} &=& \int \frac{d^3 k ~d^3 k'}{(2 \pi)^{3}} \overline{E(\vec{k}) E(\vec{k'})} \int \frac{d^2 \Omega}{4 \pi} \left[ \psi_k(\eta_s^{(0)}) e^{i r \vec{k}\cdot \hat{{x}}}\right]_{r=\eta_0 - \eta_s^{(0)}} \left[ \psi_{k'}(\eta_s^{(0)}) e^{i r  \vec{k'}\cdot\hat{{x}}} \right]_{r=\eta_0 - \eta_s^{(0)}} \nonumber \\
&=&  \int \frac{d^3 k}{(2 \pi)^{3}} ~ |\psi_k(\eta_s^{(0)})|^2 \int_{-1}^{1} \frac{d(\cos\theta)}{2} \left[ e^{i k \Delta \eta \cos\theta} \right] \left[ e^{-i k \Delta \eta \cos\theta} \right] \nonumber \\
&=&  \int \frac{d^3 k}{(2 \pi)^{3}}  ~ |\psi_k(\eta_s^{(0)})|^2 = \int_0^{\infty} \frac{d k}{k}  ~ {\cal P}_{\psi}(k,\eta_s^{(0)}),
\label{TrivialExample1}
\eea
\bea
\overline {\lla \psi_s \rra \lla \psi_s \rra} &=& \int \frac{d^3 k ~d^3 k'}{(2 \pi)^{3}} \overline{E(\vec{k}) E(\vec{k'})} \left[ \int \frac{d^2 \Omega}{4 \pi} \psi_k(\eta_s^{(0)}) e^{i r \vec{k}\cdot \hat{{x}}} \right]_{r=\eta_0 - \eta_s^{(0)}} \left[ \int \frac{d^2 \Omega'}{4 \pi} \psi_{k'}(\eta_s^{(0)}) e^{i r  \vec{k'}\cdot\hat{{x}}'} \right]_{r=\eta_0 - \eta_s^{(0)}} \nonumber \\
&=&  \int \frac{d^3 k}{(2 \pi)^{3}} ~ |\psi_k(\eta_s^{(0)})|^2 \left[ \int_{-1}^{1} \frac{d(\cos\theta)}{2} e^{i k \Delta \eta \cos\theta} \right] \left[ \int_{-1}^{1} \frac{d(\cos\theta')}{2} 
e^{-i k \Delta \eta \cos\theta'} \right] \nonumber \\
&=&  \int \frac{d^3 k}{(2 \pi)^{3}}  ~ |\psi_k(\eta_s^{(0)})|^2 \left( \frac{\sin(k \Delta\eta)}{k \Delta \eta} \right)^2 = 
 \int_0^{\infty} \frac{d k}{k}  ~ {\cal P}_{\psi}(k,\eta_s^{(0)}) \left( \frac{\sin(k \Delta\eta)}{k \Delta \eta} \right)^2,
 \label{TrivialExample1a}
\eea
where in the second line of both terms we made use of isotropy (i.e. $\psi_k$ only  dependent on $k = |\vec k|$), and defined $\theta$ and $\theta'$ as the angles between $\vec{k}$ and $\vec{x} \equiv r \hat{{x}}$ and between $\vec{k}'$ and $\vec{x}' \equiv r \hat{{x}}'$.
We recall that $\Da \eta = \eta_o-\eta_s^{(0)}$.
We have also introduced the (so-called dimensionless) power spectrum of $\psi$:
\be
{\cal P}_\psi (k, \eta)  \equiv \frac{k^3}{2 \pi^2}  |\psi_k(\eta)|^2 .
\label{DefSpectra}
\ee
To give a slightly more involved example, we can consider the average of a term like $\psi_s {Q_s}/{\Delta \eta}$:
\bea
\overline{\lla \psi_s \frac{Q_s}{\Delta \eta} \rra} &=& - \frac{2}{\Delta \eta} \int \frac{d^3 k d^3 k'}{(2 \pi)^3}\overline{E(\vec{k}) E(\vec{k}')} \int \frac{d^2 \Omega}{4 \pi} \left[ \psi_{k}(\eta_s^{(0)})  e^{i \vec{k} \cdot \hat{x} \Delta\eta} \int_{\eta_s^{(0)}}^{\eta_o} d\eta\, \psi_{k'}( \eta)  e^{i \vec{k}' \cdot \hat{x} (\eta_o - \eta)} \right] \nonumber \\
&\Rightarrow &  - 2 \int_{0}^{\infty} \frac{d k}{k} \Pcal_\psi(k, \eta_o) \frac{ {\rm SinInt}(k \Delta \eta)}{k \Delta \eta}\, ,
\label{TrivialExample2}
\eea
\bea
\overline{\lla \psi_s \rra \lla \frac{Q_s}{\Delta \eta} \rra} &=& - \frac{2}{\Delta \eta} \int \frac{d^3 k d^3 k'}{(2 \pi)^3}   \overline{E(\vec{k}) E(\vec{k}')} \left[ \int \frac{d^2 \Omega}{4 \pi} \psi_{k}(\eta_s^{(0)})  e^{i \vec{k} \cdot \hat{x} \Delta\eta} \right] \left[ \int \frac{d^2 \Omega'}{4 \pi}  \int_{\eta_s^{(0)}}^{\eta_o} d\eta \,\psi_{k'}( \eta) e^{i \vec{k}' \cdot \hat{x'} (\eta_o - \eta)} \right] \nonumber \\
&\Rightarrow& - 2  \int_{0}^{\infty} \frac{d k}{k} \Pcal_\psi(k, \eta_o)  ~ \frac{\sin(k \Delta\eta)}{k \Delta \eta}  \frac{ {\rm SinInt}(k \Delta \eta)}{k \Delta \eta}\, , 
\label{TrivialExample2a}
\eea
where the arrows refer to the case of a time-independent fluctuation mode and
\beq
 {\rm SinInt}(x) \equiv  \int_0^x {dy\over y} \sin y ~~.
\eeq
As one can see, the angular average is making the results completely different in the two cases. We remark that  the presence of the ${\rm SinInt}$ function is a direct consequence of the integration over time in $Q_s$. Consequently, the non-local nature of the backreaction terms is reflected in the form of the corresponding spectral coefficients.

Following this approach, all the relevant contributions to the averaged functions of the luminosity redshift relation, at second order, can be parameterized in the form:
\begin{eqnarray}
\overline{\langle X \rangle} &=& \int_0^{\infty} \frac{d k}{k}  ~ {\cal P}_{\psi}(k, \eta_o)  \Ccal_X(k,\eta_o,\eta_s^{(0)}),  \\
\overline{\langle X' \rangle \langle Y' \rangle} &=& \int_0^{\infty} \frac{d k}{k}  ~ {\cal P}_{\psi}(k, \eta_o)  
\Ccal_{X'}(k,\eta_o,\eta_s^{(0)}) \Ccal_{Y'}(k,\eta_o,\eta_s^{(0)}),  
\end{eqnarray}
valid for any given model of perturbation spectrum. 
Here, $X'$ and $Y'$ ($X$) are first (second) order  generic functions of $(\eta,r,\theta^a)$, and the $\Ccal$ are the associated spectral coefficients. In the particularly simple case of a CDM-dominated  geometry the spectral distribution of sub-horizon scalar  perturbations is time-independent ($\pa_\eta \psi_k=0$) and, as we shall see later, all the spectral coefficients $\Ccal$ can be  calculated analytically. 
We should stress, however, that when performing numerical calculations the above integration limits will be replaced by appropriate cut-off values determined by the physical range of validity of the considered spectrum.

%%%%%%%%%%%%%%%%%%%%%%%%%%%%%%%%%%%%%%%%%%%%%%%%%%%%%%%%%%%%%%%%%%%%%%%%%%%%%%%%%%%%%%%%%%%%%

\section{Dynamical evolution of scalar perturbations up to second-order}
\label{Sec4}
\setcounter{equation}{0}

%%%%%%%%%%%%%%%%%%%%%%%%%%%%%%%

For a full computation of the fractional correction $f_\Phi$ what we need, at this point, is the combined angular and {\em ensemble} averages of the three basic quantities, $ \Ical_{1}$, $ \Ical_{1,1}$ and $ \Ical_{2}$. We must evaluate, in particular,  the spectral coefficients $\{\Ccal_{\Tcal_i^{(1)}}\}$. $\{\Ccal_{\Tcal_i^{(1,1)}}\}$ and $\{\Ccal_{\Tcal_i^{(2)}}\}$, related to the terms defined  in Eqs. (\ref{DefI1}-\ref{DefI2}) in terms of the first and  second-order Bardeen potential. For this purpose we need to know the dynamical evolution of the scalar fluctuations, at first order for the computation of  $ \Ical_{1}$, $ \Ical_{1,1}$  and at second order for $ \Ical_{2}$.

Let us consider, first of all, a general  model with cosmological constant plus dust sources. For the evolution of the scalar degrees of freedom in the Poisson gauge we will follow the analysis performed, up to second order,  in \cite{BMR}. 
In a general $\Lambda$CDM model the linear scalar perturbation obeys the evolution equation 
\be
\psi''+3\Hcal \psi'+\left(2 \Hcal'+\Hcal^2\right) \psi=0\,.
\ee
Considering only the growing mode solution we can set
\be 
\psi(\eta,\vec x)=\frac{g(\eta)}{g(\eta_o)} \psi_o(\vec x)\, ,~~~~~~~~~~~~~~~~~~~~~~
\label{SolutionGeneralPsi}
\ee
where $g(\eta)$ is the so-called ``growth-suppression factor", or -- more precisely -- the least decaying mode solution, and $\psi_o$ is the present value of the gravitational potential. This growth factor can be expressed analytically in terms of elliptic functions \cite{Eisenstein:1997ij} (see also \cite{PeterUzan}), and it is well approximated by a simple function of the critical-density parameters of non-relativistic matter ($\Om_m$) and cosmological constant ($\Om_\La$) as follows:
\be
g=\frac{5}{2} g_{\infty}\frac{\Omega_m}{\Omega_m^{4/7}-\Omega_{\Lambda}+(1+\Omega_m/2)(1+\Omega_{\Lambda}/70)}. 
\label{gformula}
\ee
Here $g_{\infty}$ represents the value of $g(\eta)$ at early enough times when the cosmological constant was negligible, and  is fixed by the condition $g(\eta_o)=1$.

The second-order potentials obey a similar evolution equation,  containing, however, an appropriate source term.
Their final expression in terms of  $\psi_o$ has been given in \cite{BMR} and reads:
\begin{eqnarray}
\label{PSI}
\psi^{(2)}(\eta)&=&\left( 
B_1(\eta)-2g(\eta)g_{\infty} -\frac{10}{3}(a_{\rm nl}-1)g(\eta)g_{\infty}
\right)\psi_o^2 
+\left( B_2(\eta) -\frac{4}{3}g(\eta)g_{\infty}\right) \Ocal^{ij} \partial_j \psi_o \partial _i \psi_o
\nonumber \\ & &  
+ B_3(\eta)\, \Ocal_3^{ij} \partial_j \psi_o \partial _i \psi_o +B_4(\eta)\,  \Ocal_4^{ij} \partial_j \psi_o \partial _i \psi_o \, ,\\
\label{PHI}
\phi^{(2)}(\eta)&=&\left( B_1(\eta)+4g^2(\eta)
-2g(\eta)g_{\infty} -\frac{10}{3}(a_{\rm nl}-1)g(\eta)g_{\infty}
\right)\psi_o^2 
+\Bigg[ B_2(\eta)+\frac{4}{3} g^2(\eta) \left( e(\eta)+\frac{3}{2} \right)
-\frac{4}{3}g(\eta)g_{\infty} \Bigg] \nonumber \\
& & \times \Ocal^{ij} \partial_j \psi_o \partial _i \psi_o
+ B_3(\eta)\, \Ocal_3^{ij} \partial_j \psi_o \partial _i \psi_o +B_4(\eta)\,  \Ocal_4^{ij} \partial_j \psi_o \partial _i \psi_o~~,
\end{eqnarray}
where 
\beq
\label{Ocals}
\Ocal^{ij} = \nabla^{-2} \left( \delta^{ij}-3 \frac{\partial^i \partial^j}{\nabla^2} \right) ~~~~,~~~~ \Ocal_3^{ij} = \frac{\partial^i \partial^j}{\nabla^2} 
~~~~,~~~~ \Ocal_4^{ij} =\delta^{ij} ~~,
\eeq
and where we have introduced the functions $B_A(\eta)={\mathcal H}_o^{-2} \left[l(\eta_o)+3 \Omega_{m}(\eta_o)/2 \right]^{-1} \tilde{B}_A(\eta)$, with $A=1,2,3,4$, and with the following definitions:
\begin{eqnarray}
\label{B1B2}
& &\!\!\!\!\!\!\!\!\!\!\!\!\!\!\tilde{B}_1(\eta)\!= \int_{\eta_{m}}^\eta d\tilde{\eta} \,{\mathcal H}^2(\tilde{\eta}) 
(l(\tilde{\eta})-1)^2 C(\eta,\tilde{\eta})\, , \,\,\,\,
\tilde{B}_2(\eta)\!=\! 2\int_{\eta_{m}}^\eta d\tilde{\eta} \, {\mathcal H}^2(\tilde{\eta}) 
\Big[2 (l(\tilde{\eta})-1)^2-3
+3 \Omega_m(\tilde{\eta}) \Big] C(\eta,\tilde{\eta}) , \\
& &\!\!\!\!\!\!\!\!\!\!\!\!\!\!\tilde{B}_3(\eta)\!=\! \frac{4}{3} \int_{\eta_{m}}^\eta d\tilde{\eta} \left(e(\tilde{\eta})
+\frac{3}{2} \right) C(\eta,\tilde{\eta}) \, , \,\,\,\,\,\,\,\,\,\,\,\,
\tilde{B}_4(\eta)\!=\! - \int_{\eta_{m}}^\eta d\tilde{\eta} \,C(\eta,\tilde{\eta})\, ,
\label{B3tildeB4tilde}
\end{eqnarray}
and with 
\begin{equation}
C(\eta,\tilde{\eta})= g^2(\tilde{\eta}) \frac{a(\tilde{\eta})}{a(\eta_o)} 
\Big[ g(\eta){\mathcal H}(\tilde{\eta})-g(\tilde{\eta}) 
\frac{a^2(\tilde{\eta})}{a^2(\eta)} {\mathcal H}(\eta) \Big] ~~~ , ~~~ e(\eta)=l^2(\eta)/\Omega_m(\eta)~~~,~~~ l(\eta)=1+g'/(\Hcal g).
\label{labelC}
\end{equation}
Here $\eta_m$ denotes the time when full matter domination starts \cite{BMR}. Its precise value is irrelevant since the region of integration around $\eta_m$ is strongly suppressed.
Finally,  $a_{{\rm nl}}$ is the so-called non-gaussianity parameter (see \cite{BMR}),  which approaches unity  in the standard inflationary scenario. 

For further use let us now evaluate the ensemble (and angular/light-cone) average of the different operators defined in Eq.(\ref{Ocals}), when applied to $\partial_i \psi_o \partial_j \psi_o$. Considering first the {\em ensemble} average of  $\Ocal^{ij} \partial_i \psi_o \partial_j \psi_o$, and Fourier-expanding $\psi_o$, we get
 (see \cite{Marozzi:2011zb}):
\beq
\overline{\Ocal^{ij} \partial_i \psi_o \partial_j \psi_o} = 
\int \frac{d^3q ~ d^3 k}{(2 \pi)^3} \delta^{(3)}(\vec{q}) e^{i \vec{q} \cdot \hat{x} r}  ~ \psi_{|\vec{k}|} \psi^*_{|\vec{k}-\vec{q}|} \left[ -\frac{2 (\vec{k}\cdot\vec{q}) + |\vec{k}|^2}{ |\vec{q}|^2} + 3 \frac{(\vec{k}\cdot\vec{q})^2}{|\vec{q}|^4} \right] \,.
\eeq 
(from this point, and up to the end of this section, we will neglect all  suffixes ``$o$" present in terms inside the integrals).
By using the Taylor expansion of $\psi^*_{|\vec{k}-\vec{q}|}$ around $\vec{q} = 0$  we have:
\beq
\psi^*_{|\vec{k}-\vec{q}|} \simeq \psi^*_{k} - \frac{\vec{k}\cdot\vec{q}}{k} \partial_k \psi^*_{k} + \frac12 \left\{ \left( \frac{q^2}{k} - \frac{(\vec{k}\cdot\vec{q})^2}{k^3} \right) \partial_k \psi^*_{k} + \frac{(\vec{k}\cdot\vec{q})^2}{k^2} \partial_k^2 \psi^*_{k} \right\} + \Ocal(q^3),
\eeq
where $k \equiv |\vec{k}|$, $q \equiv |\vec{q}|$, and where the 
latter terms have been obtained by using the Hessian matrix $H^{ij} = \partial_{k_i} \partial_{k_j} \psi^*_k = \partial_{k_i} [({k^j}/{k}) \partial_{k} \psi^*_k ]$. Combining the  last two results, writing the integral over $k$ as $\int_0^\infty 2 \pi k^2 dk \int_{-1}^{+1} d\cos\alpha (...)$, where  $\vec{k}\cdot\vec{q} = k q \cos\alpha$, and integrating over $\cos\alpha$, we obtain:
\beq
\overline{\Ocal^{ij} \partial_i \psi_o \partial_j \psi_o} = \int \frac{d^3q}{(2 \pi)^3} \delta^{(3)}(\vec{q}) e^{i \vec{q} \cdot \hat{{x}} r } \int_0^\infty 2 \pi k^2 dk ~ \left[ \frac{16 k}{15} ~ \psi_k \partial_k \psi^*_k 
+ \frac{4 k^2}{15} ~ \psi_k \partial_k^2 \psi^*_k \right]  \,.
\label{Op}
\eeq
Note that the integrand's dependence on the angle $\theta$ between $\vec{x}$ and $\vec{q}$ only arises from the exponential term $\exp(i \vec{q} \cdot \hat{{x}} r )$, which disappears in the presence of $\delta^{(3)}(\vec{q})$. As a consequence, the angular average has no impact on this particular term and we get:
\beq
\overline{\lla \Ocal^{ij} \partial_i \psi_o \partial_j \psi_o \rra} = 
\overline{\Ocal^{ij} \partial_i \psi_o \partial_j \psi_o}=\int_0^\infty  dk \frac{k^2}{2 \pi^2} ~ \left[ \frac{8k }{15} ~ \psi_k \partial_k \psi^*_k + \frac{2 k^2}{15} ~ \psi_k \partial_k^2 \psi^*_k \right]  \,.
\label{Final_overline_O}
\eeq
Let us note also that, quite generally, $\partial_k \psi_k \sim \psi_k/k$, and thus the above term is of the same order as  
$\overline{\lla \psi_o \psi_o \rra}$.

By repeating exactly the same procedure for the $\Ocal_3^{ij} \partial_i \psi_o \partial_j \psi_o$ term, we find:
\beq
\overline{\Ocal_3^{ij} \partial_i \psi_o \partial_j \psi_o} = \int \frac{d^3q}{(2 \pi)^3} \delta^{(3)}(\vec{q}) e^{i \vec{q} \cdot \hat{{x}} r } \int_0^\infty 4 \pi k^2 dk ~ \left[ \frac{k^2}{3} 
|\psi_k|^2 + \frac{k q^2}{5} ~ \psi_k \partial_k \psi^*_k + \frac{k^2 q^2}{20} ~ \psi_k \partial_k^2 \psi^*_k \right] \,,
\label{O3p}
\eeq
where the  last  two contributions  are vanishing  because of the $q$-integration. We are thus left with the following simple result (insensitive to the angular average, as before)
\beq
\overline{\lla \Ocal_3^{ij} \partial_i \psi_o \partial_j \psi_o \rra} = 
\overline{\Ocal_3^{ij} \partial_i \psi_o \partial_j \psi_o}
=\int_0^\infty \frac{dk}{k} ~ \left[ \frac{k^2}{3} {\cal P}_\psi(k,\eta_o)\right] ~.
\label{Final_overline_O3}
\eeq
Finally,  the last term $\Ocal_4^{ij} \partial_i \psi_o \partial_j \psi_o=|\vec{\nabla} \psi_o|^2$ is trivial, and gives 
\beq
\overline{\lla \Ocal_4^{ij} \partial_i \psi_o \partial_j \psi_0 \rra} = 
\overline{\Ocal_4^{ij} \partial_i \psi_o \partial_j \psi_o}
=\int_0^\infty \frac{dk}{k} ~ \left[ k^2 {\cal P}_\psi(k,\eta_o)\right] ~.
\label{Final_overline_O4}
\eeq

%%%%%%%%%%%%%%%%%%%%%%%%%%%%%%%%%%%%%%%%%%%%%%%%%%%%%%%%%%%%%%%%%%%%%%%%%%%%%%%%%%%%%%%%%%%%%

\section{An illustrative example: the CDM model}
\label{Sec5}
\setcounter{equation}{0}
%%%%%%%%%%%%%%%%%%%%%%%%%%%%%%

In the previous sections we have computed, up to the second perturbative order, the general form of the corrections induced by the stochastic fluctuations of the geometry on the averaged luminosity flux, for a spatially flat FLRW metric and for a generic spectrum of metric perturbations. We have found that such corrections are controlled by the combined angular and {\em ensemble} averages of three basic quantities, $ \Ical_{1}$, $ \Ical_{1,1}$ and $ \Ical_{2}$. In this section we will evaluate such averages for the particular case of the standard CDM model, as a simple illustrative example. The results we will obtain, will give us useful information about  the dominant terms to be selected also for the discussion of the phenomenologically more relevant case, the $\Lambda$CDM model, presented in the forthcoming sections.

%%%%%%%%%%%%%%%%%%%%%%%%%%%%%%%%%%
\subsection{The quadratic first-order contributions $\overline{\lla \Ical_1 \rra^2}$ and $\overline{\lla \Ical_{1,1} \rra}$}
\label{Sec5A}
%%%%%%%%%%%%%%%%%%%%%%%%%%%%%%%%%%%%%%%%%%%%%%%%%%%%%%%%%%%%%%%%%%%%%%%%%%%%%%

Let us now explicitly calculate the spectral coefficients for the first two  backreaction  terms (induced by the averaging)  contributing to Eq. \rref{7a}. The genuine second-order term $\overline{\lla \Ical_2 \rra}$ will be discussed in the next subsection. We start considering the first term,  $\overline{\lla \Ical_1 \rra^2}$. From Eq.(\ref{DefI1}) we obtain: 
\beq
\overline{\lla \Ical_1 \rra^2} = \sum_{i=1}^{3} \sum_{j=1}^{3} \overline{\lla {\Tcal^{(1)}_i} \rra \lla {\Tcal^{(1)}_j} \rra} 
= \int_0^{\infty} \frac{d k}{k} ~ {\cal P}_{\psi}(k, \eta_o) ~ \sum_{i=1}^3 \sum_{j=1}^3  \Ccal_{\Tcal^{(1)}_i}(k,\eta_o,\eta_s)~ 
\Ccal_{\Tcal^{(1)}_j}(k,\eta_o,\eta_s)
\label{Eq_with_spectralI_1^2}
\eeq
(notice that, from now on, the background solution $\eta_s^{(0)}$ will be simply denoted by $\eta_s$). 
For a dust-dominated phase the spectral distribution of sub-horizon scalar perturbations is time independent $\partial_\eta \psi_k=0$, and the scale factor $a(\eta)$ can be written as $a(\eta) = a(\eta_o)(\eta/\eta_o)^2$. We can also define
\beq
f_{o,s} \equiv \int_{\eta_{in}}^{\eta_{o,s}} d\eta \frac{a(\eta)}{a(\eta_{o,s})} = \frac{\eta_{o,s}^3 - \eta_{in}^3}{3\eta_{o,s}^2} ~ \simeq \frac13 \eta_{o,s}
\label{fCDM}
\eeq
(recall that $\eta_{in}$ satisfies, by definition, $\eta_{in} \ll \eta_{o,s}$). All the spectral coefficients of Eq. (\ref{Eq_with_spectralI_1^2}) can then be easily calculated, and the result is reported in Table \ref{Tab1}~\footnote{We take the opportunity 
to point out two misprints appearing in Table 2 of \cite{BGMNV1} where the spectral coefficients of both $A_1$ and $A_2$ should 
have the opposite sign.}. 
In a similar way,  the second contribution to Eq. (\ref{7a}) can be expressed as:
\beq
\label{IntkI11}
\overline{\lla \Ical_{1,1} \rra} = \int_0^{\infty} \frac{d k}{k} ~ {\cal P}_{\psi}(k, \eta_o) ~ \sum_{i=1}^{23} \Ccal_{\Tcal_i^{(1,1)}},
\eeq
and the explicit form of the spectral coefficients $\Ccal_{\Tcal_i^{(1,1)}}$, for the CDM case, is presented in Appendix B. 
 
 \begin{table}[t]
\centering
\caption[]{\label{Tab1} The spectral coefficients $\Ccal_{\Tcal^{(1)}_i}$ for the  $\overline{\lla {\Tcal^{(1)}_i} \rra \lla {\Tcal^{(1)}_j} \rra}$ terms.}
\vskip 0.2 cm
\begin{tabular}{|c|c|}
\hline
$\Tcal^{(1)}_i$ & $ \Ccal_{\Tcal^{(1)}_i} (k,\eta_0,\eta_s)$ \\
\hline\hline
$\Tcal^{(1)}_1$ & $-2 \frac{\sin k \Da \eta}{k \Da \eta}$ \\ \hline
$\Tcal^{(1)}_2$ & $-2 \left(1 - \frac{1}{{\mathcal H}_s\Delta \eta}\right) \left(1 - \frac{\sin k \Da \eta}{k \Da \eta}\right)+2 \left(1 - \frac{1}{{\mathcal H}_s\Delta \eta}\right) \frac{f_s}{\Delta \eta}  \left(\cos k \Da \eta - \frac{\sin k \Da \eta}{k \Da \eta}\right) $ \\ \hline
$\Tcal^{(1)}_3$ & $ \frac{4}{k \Da \eta} {\rm SinInt}(k \Da \eta)$ \\ \hline
\end{tabular}
\end{table}
\vskip 0.3 cm

Considering all contributions generated by $\Ical_1$ and $\Ical_{1,1}$, we find that the dominant contributions are all contained 
in $\overline{\lla \Ical_{1,1} \rra}$, and are characterized by spectral coefficients proportional to $k^2$ (such dominant terms  have been  emphasized, in the Appendix B, by enclosing them in a rectangular box). They correspond, in particular, to  the terms $\Tcal_i^{(1,1)}$ with $\{i=2,4,5,7,8,12,14,18,20\}$. Including only such dominant contributions 
we find that, to leading order, 
\bea
\left[ \sum_{i=1}^{23}\Ccal_{\Tcal_i^{(1,1)}}\right]_{\mbox{Lead}} &=& \aleph_s \frac{f_s^2 + f_o^2}{3} k^2 + \frac{2 (\Xi_s - 3)}{\Hcal_s} \frac{f_s}{3} k^2 + \Xi_s \frac{f_s^2}{3} k^2 - \Xi_s \frac{f_o^2}{3} k^2 \nonumber \\
&=& f_o^2 k^2 \left(\frac{\aleph_s}{3} - \frac{\Xi_s}{3} \right) + f_s^2 k^2 \left(\frac{\aleph_s}{3} + \frac{4 \Xi_s}{3} - 3 \right)\nonumber \\
&=& - \frac{k^2}{\Hcal_0^2} \ti f_{1,1}(z) \,,
\eea
where we have defined 
\beq
\aleph_s = \Xi_s^2 - \frac{1}{\Hcal_s \Delta\eta}\left(1 - \frac{\Hcal_s'}{\Hcal_s^2}\right),
\eeq
and we have used the relation $\Hcal_s \simeq 2 / (3 f_s)$. Also, we have included into the function  $\ti f_{1,1}(z)$ all the $z$-dependence of these leading contributions. After some simple algebra we  find: 
\begin{equation}
\ti f_{1,1}(z) =  \frac{10-12 \sqrt{1+z}+5 z \left(2+\sqrt{1+z}\right)}{27\, (1+z) \left(-1+\sqrt{1+z}\right)^2} \,. 
\label{f_11}
\end{equation}
This function (and thus the corresponding backreaction) flips sign  around  $z^{\ast}=0.205$ (as illustrated also in Fig. \ref{f1}). 

It should be noted, finally, that some of the genuine second-order terms contained into $\Ical_2$ are also associated to spectral coefficients proportional to $k^2$. However, as we shall see in the next subsection, the corresponding contributions to Eq. \rref{7a} turn out to be roughly an order of magnitude smaller than the above ones,   because of approximate cancellations.

%%%%%%%%%%%%%%%%%%%%%%%%%%%%%%%%%%%%%%%%%%%%%%%%%%%%%%%%%%%%%%%%%%%%%%
\subsection{The genuine second-order contribution $\overline{\lla \Ical_2 \rra}$}
\label{Sec5B}
%%%%%%%%%%%%%%%%%%%%%%%%%%%%%%%%%%%%%%%%%%%%%%%%%%%%%%%%%%%%%%%%%%%%%%%%

In order to complete the calculation of $f_\Phi(z)$ we still have to consider the genuine second-order backreaction term  $\overline{\lla \Ical_2 \rra}$. In particular, we  must evaluate the spectral coefficients $\{\Ccal_{\Tcal_i^{(2)}}\}$, corresponding to the various contributions defined  in Eq. \rref{DefI2}, in terms of the first-order Bardeen potential. By using the results of Sect. \ref{Sec4}, and starting from Eqs. (\ref{PSI}) and (\ref{PHI}), we can easily see that the only possible  $k^2$-enhanced contributions arising from the coefficients  $\{\Ccal_{\Tcal_i^{(2)}}\}$ (which  only contain functions of $\phi^{(2)}$ and $\psi^{(2)}$) should correspond to the term $B_3(\eta) \nabla^{-2} \partial_i\partial^j(\partial^i \psi_o \partial_j \psi_o )+B_4(\eta) \partial^i \psi_o \partial _i\psi_o$. Therefore, we should obtain  $\phi^{(2)}\simeq \psi^{(2)}$ at leading order.

Let us discuss and estimate all possible contributions  for the CDM model we are considering in this section.  In this simple case  
$l(\eta)=1$ and  $B_1(\eta)=B_2(\eta)=0$ (see Eq.(\ref{B1B2})). Furthermore, restricting our attention to the standard inflationary scenario, we can set $a_{nl}=1$. Eqs.(\ref{PSI}) and (\ref{PHI}) thus reduce to: 
\bea
\psi^{(2)} &=&  -2\psi_o^2 - \frac{4}{3} ~ \Ocal^{ij} \partial_i \psi_o \partial_j \psi_o + B_3(\eta) ~ \Ocal_3^{ij} \partial_i \psi_o \partial_j \psi_o
+ B_4(\eta) ~ \Ocal_4^{ij} \partial_i \psi_o \partial_j \psi_o \,, 
\label{412}
\\
\phi^{(2)} &=& 2\psi_o^2 + 2 ~ \Ocal^{ij} \partial_i \psi_o \partial_j \psi_o + B_3(\eta) ~ \Ocal_3^{ij} \partial_i \psi_o \partial_j \psi_o
+ B_4(\eta) ~ \Ocal_4^{ij} \partial_i \psi_o \partial_j \psi_o \,. 
\label{412a}
\eea
Finally, we can use (as before) the  scale factor  $a(\eta) = a(\eta_o) \left({\eta}/{\eta_o}\right)^2$, and obtain (according to Eq. (\ref{B3tildeB4tilde})): 
\be
B_3(\eta)=\frac{20}{21}\frac{1}{ {\cal H}^2}  ~~~~,~~~~  B_4(\eta)=-\frac{2}{7}\frac{1}{{\cal H}^2} \,.
\ee

We are now in the position of evaluating the genuine second-order terms, by exploiting the results given in  Eqs.(\ref{Final_overline_O}), (\ref{Final_overline_O3}) and (\ref{Final_overline_O4}). Following the  classification of Eq.(\ref{Tcal2}) we can see that the first two terms $\Ccal(\Tcal_1^{(2)}) $ and $\Ccal(\Tcal_2^{(2)})$ exactly cancel for the CDM case (while they cancel only at leading order for the case of a $\Lambda$CDM model). For CDM we have, in particular, 
\be
\Ccal(\Tcal_1^{(2)})= -\Ccal(\Tcal_2^{(2)})= - \frac{\Xi_s}{252} (\eta_s^2 - \eta_o^2) k^2 \,.
\ee

Another interesting simplification concerns the terms $\Tcal_{3}^{(2)}$, $\Tcal_{4}^{(2)}$ and $\Tcal_{5}^{(2)}$, namely those terms for which the integrand contains $\partial_r\psi^{(2)}$ or $\pa_r \phi^{(2)}$. 
From our previous results, in particular from Eqs. (\ref{Final_overline_O}), (\ref{Final_overline_O3}) and (\ref{Final_overline_O4}), it is easy to see that the $\psi^{(2)}$ and $\phi^{(2)}$ contributions are 
unchanged when one averages over the 2-sphere (i.e. over $\theta$ by isotropy), since the $\theta$-dependence is removed by the presence of $\delta^{(3)}(\vec{q})$. On the other hand, the presence of the $r$-derivative brings a further factor $|\vec{q}| \cos \theta$, and the $q$-integration gives zero, even before performing the angular average. 
It follows that, for a general model,
\beq
\Ccal(\Tcal_{3}^{(2)}) = 0 ~~;~~~~~~~~~ \Ccal(\Tcal_{4}^{(2)}) = 0 ~~.
\eeq
The contribution of $\Tcal_{5}^{(2)}$, on the contrary, is nonvanishing because of the presence of the partial derivative $\partial_\eta$, acting on the $B_A(\eta)$ coefficients. For the CDM model we find, in particular,
\beq
\Ccal(\Tcal_{5}^{(2)}) = \frac{\Xi_s}{126} (\eta_s^2 - \eta_o^2) k^2. 
\eeq
Finally, for the last two terms $\Tcal_{6}^{(2)}$ and $\Tcal_{7}^{(2)}$ we obtain:
\bea
\Ccal(\Tcal_{6}^{(2)}) &=& -2 - \frac{k^2 \eta_s^2}{126} + \frac{32 k}{45}
\frac{\psi_k \partial_k \psi_k^\ast}{|\psi_k|^2} + \frac{8 k^2}{45}
\frac{\psi_k \partial_k^2 \psi_k^\ast}{|\psi_k|^2} ~~, \nonumber \\
\Ccal(\Tcal_{7}^{(2)}) &=& \frac{\eta_o^3 - \eta_s^3}{189 \Delta\eta}k^2 +
\frac{16 k}{45} \frac{\psi_k \partial_k \psi_k^\ast}{|\psi_k|^2} + \frac{4
k^2}{45} \frac{\psi_k \partial_k^2 \psi_k^\ast}{|\psi_k|^2} ~~.
\eea

The sum of all contributions  then leads to:
\beq
\label{GenSecOrder}
\sum_{i = 1}^{7} \Ccal(\Tcal_i^{(2)}) = - 2 + {1\over 7}\left[ \frac{\Xi_s}{2}
(f_s^2 - f_o^2) - \frac{f_s^2}{2} - \frac{f_s^3 - f_o^3}{ \Delta \eta}
\right] k^2 + \frac{16 k}{15} \frac{\psi_k \partial_k
\psi_k^\ast}{|\psi_k|^2} + \frac{4 k^2}{15} \frac{\psi_k \partial_k^2
\psi_k^\ast}{|\psi_k|^2} ~~.
\eeq
All the above spectral coefficients are now to be numerically evaluated by using the power spectrum of the CDM model. We can easily check, however, that the leading $k^2$-contributions of these coefficients are given by:
\beq
\left[ \sum_{i = 1}^{7} \Ccal(\Tcal_i^{(2)}) \right]_{\mbox{Lead}} = {1\over 7}\left[ \frac{\Xi_s}{2} (f_s^2 - f_o^2) - \frac{f_s^2}{2} - \frac{f_s^3 - f_o^3}{\Delta \eta} \right] k^2 = - \frac{k^2}{{\cal H}_o^2} \ti{f}_2(z),
\eeq
where:
\beq
\ti{f}_2(z)= -\frac{1}{189} \frac{2-2 \sqrt{1+z}+z \left(9-2 \sqrt{1+z}\right)}{(1+z)(\sqrt{1+z}-1)}.
\label{f_2}
\eeq
Such second-order contributions turn out to be about one order of magnitude smaller than the leading contributions of the squared first-order terms of Sect. \ref{Sec5A} (as can be easily checked, for instance, by comparing the plots of   $\tilde{f}_{1,1}$ and $\tilde{f}_{2}$).

%%%%%%%%%%%%%%%%%%%%%%%%%%%%%%%%%%%%%%%%%%%%%%%%%%%%%%%%%%%%%%%%%%%%%
\subsection{Full numerical results for the CDM model}
\label{Sec5C}
%%%%%%%%%%%%%%%%%%%%%%%%%%%%%%%%%%%%%%%%%%%%%%%%%%%%%%%%%%%%%%%%%%%%%%%%%%%%%%%

At this point, in order to perform the numerical computations, we need to insert the explicit form of the power spectrum. Limiting ourselves to sub-horizon perturbations we can simply obtain $\psi_k$, for the CDM model, by applying an appropriate, time-independent transfer function to the primordial (inflationary) spectral distribution (see e.g. \cite{DurCos}). 

The power spectrum of the Bardeen potential is then given by:
\be
\label{PsiP}
{\cal P}_\psi (k)  = \left(\frac{3}{5}\right)^2 \Delta_{\cal R}^2 T^2(k) ~ , ~~~~~~~~~~~~~ \Delta_{\cal R}^2=A \left(\frac{k}{k_0}\right)^{n_s-1} ~,
\ee
where $T(k)$ is the  transfer function  which takes into account the sub-horizon evolution of the modes re-entering during the radiation-dominated era, and  $\Delta_{\cal R}^2$ is the primordial power spectrum of  curvature perturbations outside the horizon. The typical parameters of such a spectrum, namely the amplitude $A$,  the spectral index $n_s$ and the scale $k_0$, are determined by the results of recent WMAP observations  \cite{WMAP7}. In our computations we will use, in particular, the following approximate values:
\beq
 A=2.41 \times 10^{-9} ~, ~~~~~~~~  n_s=0.96 ~, ~~~~~~~~ k_0 = 0.002 \,{\rm Mpc}^{-1} ~.
\eeq
Finally, since our main purpose here is to present an illustrative example, 
it will be enough for our needs  to approximate $T(k)$ by the effective shape of the transfer function for density perturbations without  baryons, namely $T(k)=T_0(k)$, where \cite{Eisenstein:1997ik}:
\beq
\label{EH97}
T_0(q) = \frac{L_0}{L_0+q^2C_0(q)}  ~, ~~~~ L_0(q) = \ln(2e+1.8q) ~, 
~~~~ C_0(q) = 14.2+\frac{731}{1+62.5q} ~, ~~~~q  = \frac{k}{13.41 k_{\rm eq}} ~,
\eeq
and where $k_{\rm eq}\simeq 0.07 \,\Omega_{m0} h^2 Mpc^{-1}$ is the scale corresponding to matter-radiation equality, with $h\equiv H_0/(100 \,{\rm km \,s}^{-1} {\rm Mpc}^{-1})$.  

We can easily check that the above transfer function goes to 1 for $k \ll k_{\rm eq}$,  while it falls like  $k^{-2}\log k$ for $k \gg k_{\rm eq}$. 
For the numerical estimates  we will use $h=0.7$ and we will set  $a_o =1$, $\Omega_m =1$. In that case we obtain  
$k_{\rm eq}\simeq 0.036 \,{\rm Mpc}^{-1}$ (see \cite{Eisenstein:1997ik}), and we can more precisely define the asymptotic regimes of our transfer function as $T_0 \simeq 1$ for $k \laq 10^{-3}\, {\rm Mpc}^{-1}$, and $T_0 \sim k^{-2}\log k$ for  $k \gaq 2.5\, {\rm Mpc}^{-1}$.

Following the results of the previous subsections we can now set
\beq
f_\Phi(z) = \int_0^{\infty} \frac{d k}{k} \, {\cal P}_\psi(k, z=0) \Big[f_{1,1}(k,z) +  f_{2}(k,z) \Big] , 
\label{9}
\eeq
where $f_{1,1}$ and  $f_{2}$ are complicated --but known for the CDM case-- analytic functions of their arguments. 
However, as already stressed, the leading contributions in the range of $z$ relevant to  
dark-energy phenomenology are sourced by terms of the type $f(k, z) \sim (k/ {\cal H}_o)^2 \ti f(z)$. In that range of $z$ we can thus write, to a very good accuracy:
\beq
f_\Phi(z) \simeq \Big[\ti f_{1,1}(z) +  \ti f_{2}(z) \Big] \int_0^{\infty} \frac{d k}{k} \,\left(  \frac{k}{  {\cal H}_o}\right)^2{\cal P}_\psi(k, z=0)  , 
\label{10}
\eeq 
where $\ti f_{1,1}(z)$ and $\ti f_{2}(z)$ are given, respectively, by Eqs. (\ref{f_11}) and (\ref{f_2}). 

\begin{figure}[t!]
\centering
\includegraphics[width=12cm]{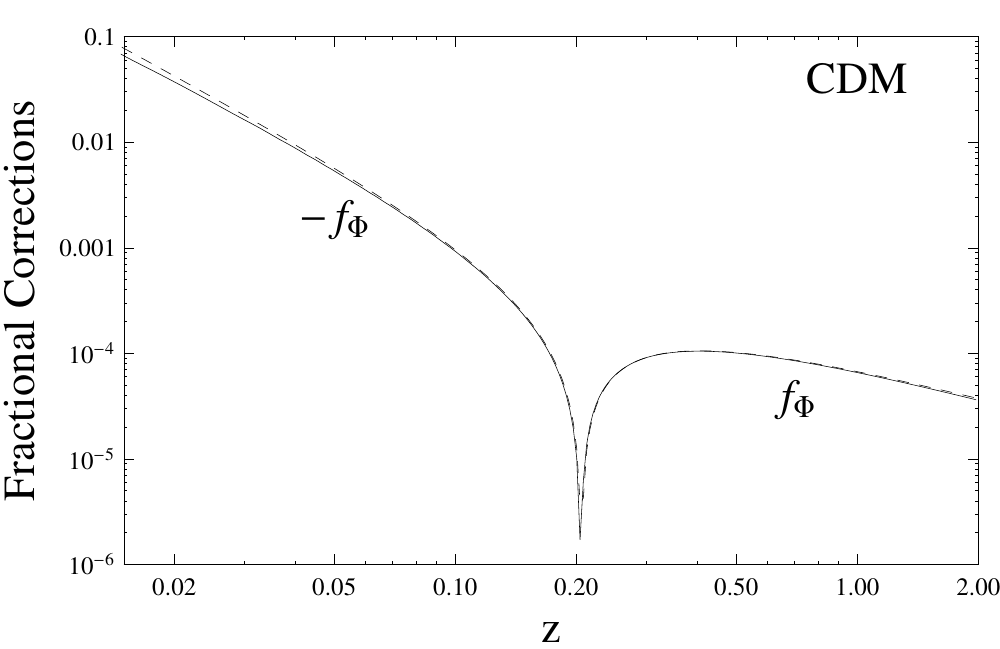}
\centering
\caption{The fractional correction $f_\Phi$ of Eq. (\ref{9}) (solid curve), compared with the same quantity given to leading order by Eq. (\ref{10}) (dashed curve), in the context of an inhomogeneous CDM model. We have used for the spectrum the one defined in Eq. (\ref{PsiP}).  The plotted curves refer, as an illustrative example, to an UV cutoff $k_{UV}=1 {\rm Mpc}^{-1}$. }
 \label{f1}
\end{figure}

To proceed, we need to insert  a power spectrum as well as infrared (IR) and ultraviolet (UV) cutoffs in (\ref{9}). The former can be identified with the present horizon ${\cal H}_o^{-1}$; however, considering the used spectra, larger scales give a completely negligible contribution. 
On the other hand, in spite of the fact that our expressions converge in the UV for any reasonable power spectrum, some mild sensitivity to the actual UV cutoff will be shown to occur in certain observables.
The absolute value (and sign) of $f_\Phi(z)$ for the CDM model, obtained  from both Eq. (\ref{9})  and (\ref{10}), are illustrated in Fig. \ref{f1}, where we can explicitly check  the accuracy of the leading order terms (\ref{10}).  
The figure also confirms that the backreaction of a realistic  spectrum of stochastic perturbations induces  negligible corrections to the averaged flux at large $z$ (the larger corrections at small $z$, due to ``Doppler terms", have been already  discussed also in \cite{BGMNV1}). 
In addition, it shows that such corrections  have the wrong $z$-dependence (in particular, they change sign at some $z$) for simulating even a tiny  dark-energy component. 

%%%%%%%%%%%%%%%%%%%%%%%%%%%%%%%%%%%%%%%%%%%%%%%%%%%%%%%%%%%%%%%%%%%%%%%%%%%%%%

%%%%%%%%%%%%%%%%%%%%%%%%%%%%%%%%%%%%%%%%%%%%%%%%%%%%%%%%%%%%%%%%%%%%%%%%%%%%%%%%%%%%%%%%%%%%%%%%%%
\section{The $\Lambda$CDM model:  power spectrum in the linear regime}
\label{Sec6}
\setcounter{equation}{0}
%%%%%%%%%%%%%%%%%%%%%%%%%%%%%%%%%%%%%%%%%%%%%%%%%%%%%%%%%%%%%%%%%%%%%%%%%%%%%%%%%%%%%%%%%%%%%%%%%%

We will now extend the procedure of the previous section to the case of the so-called ``concordance" cosmological model, using first  a  power spectrum computed in the linear  regime, and then adding the effects of  non-linearities following the parametrizations proposed in \cite{Smith:2002dz} and \cite{Takahashi}. In both cases, we will restrict our attention only to the $k^2$-enhanced terms already identified in the CDM model for the light-cone average of the flux variable, as well as to the $k^3$-enhanced terms which, as we will see, will appear in the variance, or in the averages of other functions of $d_L(z)$.

 Hereafter in all numerical computations we will use, in particular, the following numerical values: $\Omega_{\Lambda 0}=0.73$, $\Omega_{m0}=0.27$, $\Omega_{b0}=0.046$, and $h=0.7$.

%%%%%%%%%%%%%%%%%%%%%%%%%%%%%%%%%%%%%%%%%%%%%%%%%%%%%%%%%%%%%%%%%%
\subsection{Second-order corrections to the averaged luminosity flux}
\label{Sec6A}
%%%%%%%%%%%%%%%%%%%%%%%%%%%%%%%%%%%%%%%%%%%%%%%%%%%%%%%%%%%%%%%%%%

The power spectrum of the $\Lambda$CDM model is, in general, time-dependent. Considering for the moment only the linear regime, the scalar power spectrum can be written, starting from Eq.(\ref{SolutionGeneralPsi}), as 
\be
\label{PsiPpar}
\Pcal_\psi (k,\eta)=\left[\frac{g(\eta)}{g(\eta_o)}\right]^2 \Pcal_\psi (k,\eta=\eta_o) \,,
\ee
and in this case we can easily extend the  results previously obtained for the CDM model, concerning the leading ($k^2$-enhanced) contributions to the averaged-flux integral $I_\phi$. Using the general definitions of the $\Tcal_i^{(1,1)}$ terms, we first notice that such enhanced contributions arise from the following particular (sub)-terms appearing in Eq. (\ref{DefI11part2}):
\beq
\label{DefI11_Leading}
\Tcal_{2,L}^{(1,1)} = \Xi_s \left( ([\partial_r P]_s)^2 - ([\partial_r P]_o)^2 \right);
\nonumber 
\eeq
\beq
\Tcal_{4,L}^{(1,1)} = \frac12 \Xi_s (\gamma_0^{ab})_s \left( 2 \partial_a P_s \partial_b P_s \right)
~~;~~~~~~~~~~~~~~~~~~~~~~~~~~
\Tcal_{5,L}^{(1,1)} = - \Xi_s \lim_{r\rightarrow 0} \left[\gamma_0^{ab} \partial_a P  \partial_b P \right] 
~~;~~ \nonumber 
\eeq
\beq
\Tcal_{7,L}^{(1,1)} = -2 \Xi_s ([\partial_r P]_s)^2
~~;~~~~~~~~
\Tcal_{8,L}^{(1,1)} = 2 \Xi_s \frac{1}{{\mathcal H}_s} \left(\psi_s-2\int_{\eta_s}^{\eta_o} d\eta' \partial_r \psi(\eta',\eta_o-\eta', 
\tilde{\theta}^a)\right)    [\partial_r^2 P]_s~~; \nonumber 
\eeq
\beq
\Tcal_{12,L}^{(1,1)} = \left[ \Xi_s^2 - \frac{1}{{\mathcal H}_s \Delta \eta} \left( 1 - \frac{{\mathcal H}_s'}{{\mathcal H}_s^2} \right) \right] 
\left( ([\partial_r P]_s)^2 + ([\partial_r P]_o)^2 \right)
~~;~~~~~
\Tcal_{14,L}^{(1,1)} = 2 \Xi_s \left([\partial_r P]_o\right)^2
~~;~~ \nonumber 
\eeq
\beq
\Tcal_{18,L}^{(1,1)} = - \frac{1}{\Hcal_s} (\gamma_0^{ab})_s \partial_a Q_s \partial_b [\partial_r P]_s 
~~;~~~~~~~~~~~~~~~~~~~~~~~~~~~~~~~~~~~~
\Tcal_{20,L}^{(1,1)} = -2 \frac{1}{{\mathcal H}_s} [\partial_r P]_s  [\partial_r \psi]_s \, ,
\eeq
where we have added the suffix ``$L$" to stress that we are reporting here only the leading terms. 

In order to calculate the corresponding spectral coefficients, we note that their time dependence can be factorized with respect to 
the $k$-dependence whenever the time variable does not appear in the exponential factor $\exp(i \vec{k} \cdot \vec{x})$ present in our integrals (see for instance Eqs. (\ref{TrivialExample1}), (\ref{TrivialExample1a}), (\ref{TrivialExample2}), (\ref{TrivialExample2a})). This is indeed the case for the terms $\Tcal_{2,L}^{(1,1)}$, $\Tcal_{4,L}^{(1,1)}$, $\Tcal_{5,L}^{(1,1)}$, $\Tcal_{7,L}^{(1,1)}$, $\Tcal_{12,L}^{(1,1)}$, $\Tcal_{14,L}^{(1,1)}$ and $\Tcal_{20,L}^{(1,1)}$. In that case the previous CDM results for the leading  spectral coefficients can be simply generalized to the $\Lambda$CDM case through the following procedure: $(i)$ by inserting a factor $g(\eta_s)/g(\eta_o)$ whenever $\psi_s$  is present in the initial term; and $(ii)$ by replacing the $f_{o,s}$ factors  (see Eq.(\ref{fCDM})), arising from the presence of $P_{o,s}$ terms, by:
\beq
\ti{f}_{o,s} \equiv \int_{\eta_{in}}^{\eta_{o,s}} d\eta \frac{a(\eta)}{a(\eta_{o,s})}  \frac{g(\eta)}{g(\eta_{o})} .
\label{ftildeLambdaCDM}
\eeq

For  the remaining two terms $\Tcal_{8,L}^{(1,1)}$ and $\Tcal_{18,L}^{(1,1)}$ the integrals are performed along the path $r=\eta_o-\eta$, and 
the time dependence cannot be fully factorized. Therefore, the evaluation of the double integrals over $\eta$ and $k$ is much more involved than in the previous cases. However, a good approximation of the exact result can be obtained by replacing  $\psi(\eta',\eta_o-\eta', \tilde{\theta}^a)$, appearing in the integrands of $\Tcal_{8,L}^{(1,1)}$ and $\Tcal_{18,L}^{(1,1)}$, with $\psi(\eta_s,\eta_o-\eta', \tilde{\theta}^a)$ (this is so since the leading contributions to the time integral arise from  a range of values of $\eta$ approaching $\eta_s$). By adopting  such an approximation we can follow the same procedure as before, and we obtain:
\be 
\Ccal(\Tcal_{8,L}^{(1,1)}) = \frac{2}{3} \Xi_s \frac{\ti{f}_s}{\Hcal_s} k^2 \,, 
~~~~~~~~~~~
\Ccal(\Tcal_{18,L}^{(1,1)}) = - \frac{4}{3} \frac{\ti{f}_s}{\Hcal_s} k^2 \,.
\ee
This is formally the same result as in the CDM case (see Appendix B for the leading terms of $\Tcal_{8,L}^{(1,1)}$ and $\Tcal_{18,L}^{(1,1)}$), with the only difference that ${f}_s$ is replaced by $\ti{f}_s$.

Let us now move to the evaluation of the leading contributions present in the genuine second-order part $\Ical_2$. The final results of Sect. \ref{Sec5}, for the particular case of a CDM model, can be easily generalized to the $\Lambda$CDM case starting from the observation that the leading contributions can only arise  from terms containing the operators  ${\cal O}_3^{ij}$ and ${\cal O}_4^{ij}$ in Eqs.(\ref{PSI}) and (\ref{PHI}). 
As a consequence, the first two terms $\Tcal_1^{(2)}$ and $\Tcal_2^{(2)}$ will give a subleading overall contribution, while the general result that the terms $\Tcal_{3}^{(2)}$ and $\Tcal_{4}^{(2)}$ give identically zero still holds.
The remaining leading contributions can be easily obtained, using  the results in Eqs. (\ref{Final_overline_O3}) and (\ref{Final_overline_O4}), as follows:
\begin{eqnarray}
\Ccal(\Tcal_{5,L}^{(2)}) &=& -\Xi_s\left[ \frac{1}{3}\left(B_3(\eta_o)-B_3(\eta_s)\right)+\left(B_4(\eta_o)-B_4(\eta_s)\right)\right] k^2,
\\ 
\Ccal(\Tcal_{6,L}^{(2)}) &=& -\left( \frac{1}{3} B_3(\eta_s) + B_4(\eta_s)\right) k^2,
\\ 
\Ccal(\Tcal_{7,L}^{(2)}) &=& \frac{2}{\Delta\eta} \int_{\eta_s}^{\eta_o} d\eta' \left( \frac{1}{3} B_3(\eta') + B_4(\eta')\right) k^2.
\end{eqnarray}
These leading contributions can be now evaluated using Eqs. (\ref{B3tildeB4tilde}) and (\ref{labelC}) and moving to redshift space, where we can write:
\be 
\Hcal (z)=\frac{\Hcal_o}{1+z} \left[\Omega_{m0}(1+z)^3+\Omega_{\Lambda0}\right]^{1/2}\,\,\,\,\,\,,\,\,\,\,\,\,
\Omega_m=\frac{\Omega_{m0}(1+z)^3}{\Omega_{m0}(1+z)^3+\Omega_{\Lambda0}} \,\,\,\,\,\,,\,\,\,\,\,\,\Omega_{\Lambda}=\frac{\Omega_{\Lambda0}}{\Omega_{m0}(1+z)^3+\Omega_{\Lambda0}}\,,
\ee
where the suffix ``$0$" appended to $\Om_m$ and $\Om_\La$ 
denotes the present value of those fractions of critical density. We note that, as in  the CDM case,  the leading genuine second-order contributions are about one to two orders of magnitude smaller than the leading squared first-order contributions, evaluated above. 

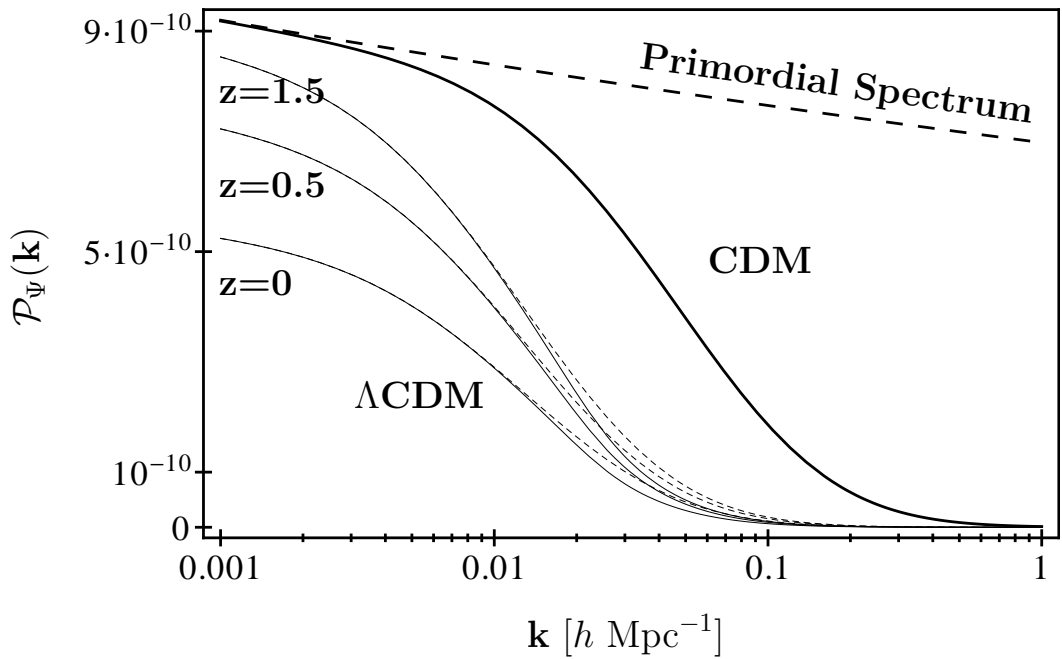
\begin{figure}[t]
\centering
\begin{tikzpicture}
\node (label) at (0,0){
        \includegraphics[width=13cm]{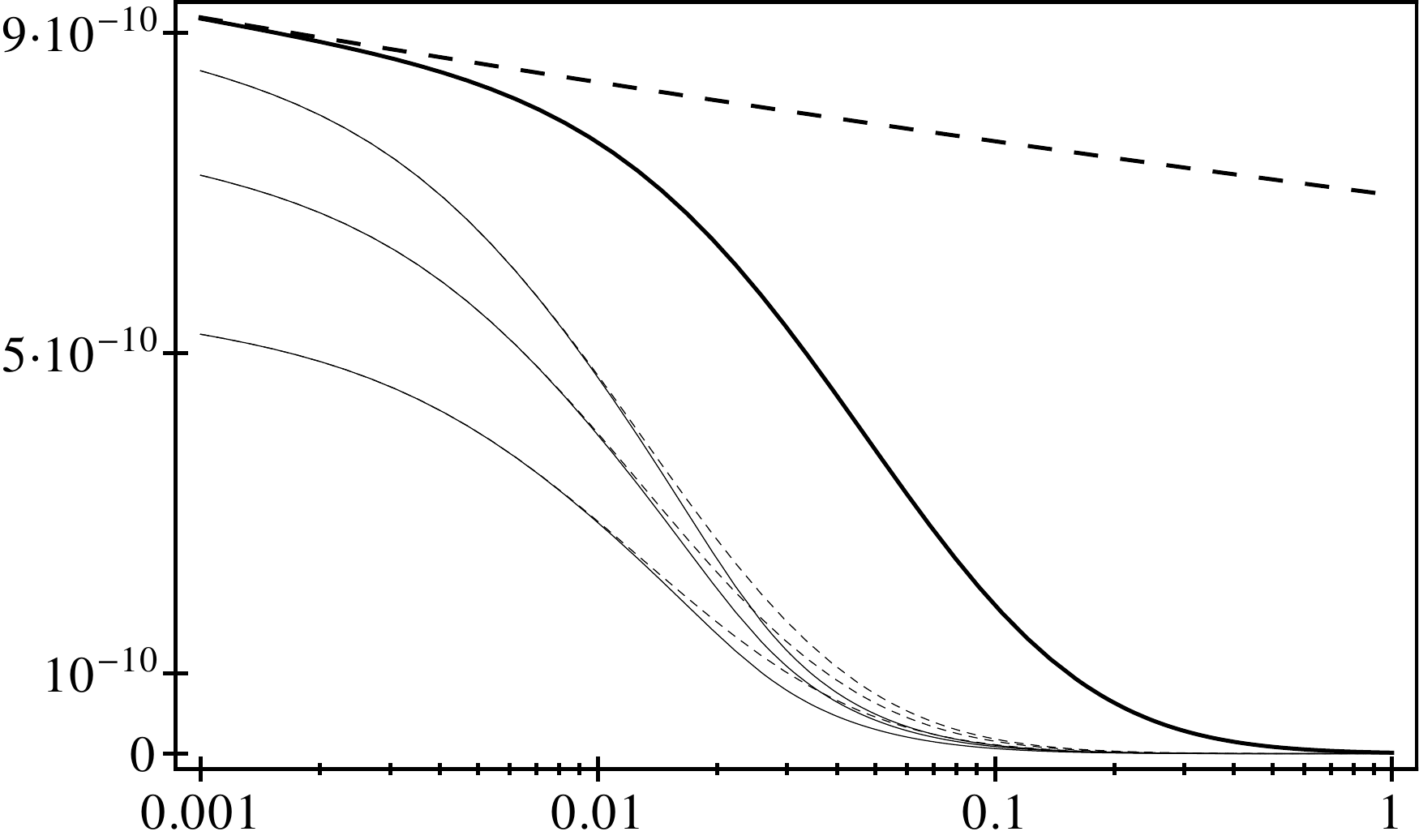}};
      
\node[anchor=west] at (-3,-1.3) (description) {\Large \textbf{$\Lambda$CDM}};
\node[anchor=west] at (1.7,0.5) (description) {\Large \textbf{CDM}};
\node[anchor=west] at (-4.8,0.2) (description) {\Large \textbf{z=0}};
\node[anchor=west] at (-4.8,1.5) (description) {\Large \textbf{z=0.5}};
\node[anchor=west] at (-4.8,2.75) (description) {\Large \textbf{z=1.5}};

\node[anchor=west,rotate=-8] at (0.8,3.2) (description) {\Large \textbf{Primordial Spectrum}};

\node[anchor=west,rotate=90] at (-7.2,-0.5) (description) {\Large $\Pcal_{\Psi}(\textbf{k})$};

\node[anchor=west] at (-0.7,-4.5) (description) {\Large $\textbf{k} ~[h ~{\rm Mpc}^{-1}]$};
\end{tikzpicture}
\caption{A comparison of the primordial inflationary spectrum (long-dashed curve) with the spectrum of the CDM model neglecting baryons (thick solid curve) and of a $\Lambda$CDM model (thin solid curves), at various values of $z$. The dotted curves for the  $\La$CDM case describe the spectrum obtained by neglecting the baryon contribution (hence without taking into account the Silk-damping effect).}
\label{Fig2}
\end{figure}

We need now to insert the explicit form of the power spectrum.
Considering the general solution of Eq. (\ref{SolutionGeneralPsi})
we can also re-express the $z$-dependence of the  power spectrum, in the linear regime, as follows:
\be
\label{PsiPz}
\Pcal_\psi (k,z)  = \left(\frac{3}{5}\right)^2 \Delta_{\cal R}^2 T^2(k) \left(\frac{g(z)}{g_{\infty}}\right)^2 ~,
\ee
where the previous CDM result (\ref{PsiP}), based on the transfer function $T(k)$ given in \cite{Eisenstein:1997ik}, is modified by the presence of the  factor $g(z)/g_\infty$ originating from the time dependence of the gravitational perturbations. 
Another modification with respect to the CDM result, implicitly contained into the transfer function $T(k)$, concerns the different numerical value of the equilibrium scale $k_{\rm eq}$ (which now turns out to be lower because of the lower value of $\Omega_{m0}$). The effects of such modifications are illustrated in Fig. \ref{Fig2} by comparing the $\Lambda$CDM  spectrum of 
Eq. (\ref{PsiPz}), at different values of $z$, with two $z$-independent spectra: the primordial spectrum of scalar perturbations $(3/5)^2  \Delta_{\cal R}^2$, and the ``transferred" spectrum of the CDM model, introduced in Sect. \ref{Sec5} (notice that, as expected, the $\Lambda$CDM and CDM spectra tend to coincide at large enough values of $z$).  In the $\La$CDM case the solid curves are obtained with a transfer function which takes into account the presence of baryonic  matter, while the dashed curves correspond to the transfer function $T(k)=T_0(k)$ of Sect.\ref{Sec5C} (without baryons). Here, however, we are  
always using a spectrum  evaluated in the linear regime.

As illustrated in Fig. \ref{Fig2}, the effect of neglecting the baryonic fraction of $\Omega_m$ (and thus the associated Silk-damping effect) may lead to an overestimation up to 40\%  of the corresponding transfer function for scalar perturbations, in the range $k \gaq 0.01 h \,{\rm Mpc}^{-1}$ (see  \cite{Eisenstein:1997ik}).  In order to take into account this baryonic contribution we have to replace the value of $q$ used in the previous section, i.e. $q  = k/(13.41 k_{\rm eq})$, with the more accurate value given by  \cite{Eisenstein:1997ik}:
\bea
q&=&\frac{k \times Mpc}{h \Gamma},\\
\Gamma&=&\Omega_{m0}h\left(\alpha_{\Gamma}+\frac{1-\alpha_{\Gamma}}{1+(0.43 k s)^4}\right),\\
\alpha_{\Gamma}&=&1-0.328\ln(431 \Omega_{m0}h^2)\frac{\Omega_{b0}}{\Omega_{m0}}+0.38 \ln(22.3 \Omega_{m0} h^2)\left(\frac{\Omega_{b0}}{\Omega_{m0}}\right)^2,\\
s&=&\frac{44.5 \ln(9.83/\Omega_{m0}h^2)}{\sqrt{1+10(\Omega_{b0}h^2)^{3/4}}} \,{\rm Mpc}.
\eea
Here $\Omega_{b0}$ is the baryon density parameter, $s$ is the sound horizon and $\Gamma$ is the $k$-dependent effective shape parameter.

We have compared the above transfer function to the one which includes baryon acoustic oscillations (BAO) \cite{Eisenstein:1997ik}, and to a transfer function calculated numerically by using the so-called ``code for anisotropies in the microwave background" (CAMB) \cite{Camb}. 
We have checked, in particular,  that the above simple form of transfer function is accurate to within a few percent compared to the one calculated numerically by CAMB, for all scales of interest. In addition, the effect of including BAO only produces oscillations of the spectrum around the above value. Since we are considering here integrals over a large range of $k$, the presence of BAO has a negligible effect on our final results, and will be neglected in the rest of the paper. 

We are now in the position of computing the fractional corrections to the averaged flux variable in a perturbed  $\Lambda$CDM geometry. As discussed before, there are complicated average integrals which can be performed only by using some approximations. Once this is done, the remaining integration over $k$ can be done numerically,  exactly as in the case of the CDM model.

In a $\Lambda$CDM context we may generally expect smaller corrections to the averaged flux, due to the fact that the perturbation spectrum $\Pcal_\psi$ is  suppressed by the presence of $g(z)$. In addition (and as already stressed) the transfer function  \cite{Eisenstein:1997ik} turns out to be suppressed, at large $k$, because of a smaller value of the parameter  $k_{\rm eq}$ (see Eq. (\ref{EH97})). These expectations are fully confirmed by an explicit numerical computation of $|f_\Phi|$, which we have performed with and without the inclusion of the baryon contributions into the transfer function $T(k)$. 

\begin{figure}[t]
\centering
\includegraphics[width=8.2cm]{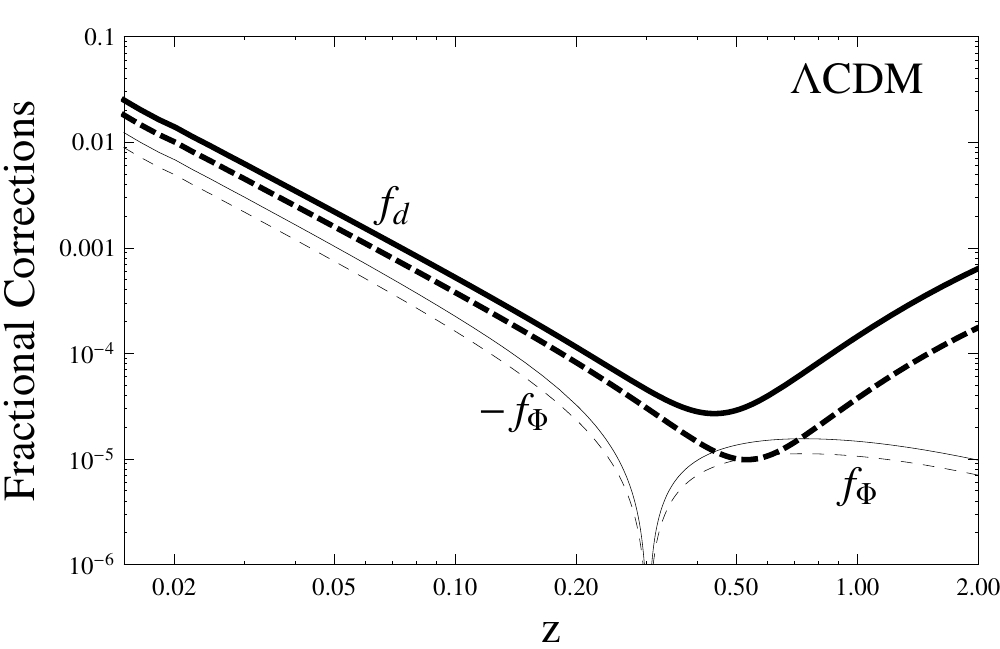}~~~~~~~~~
\includegraphics[width=8.2cm]{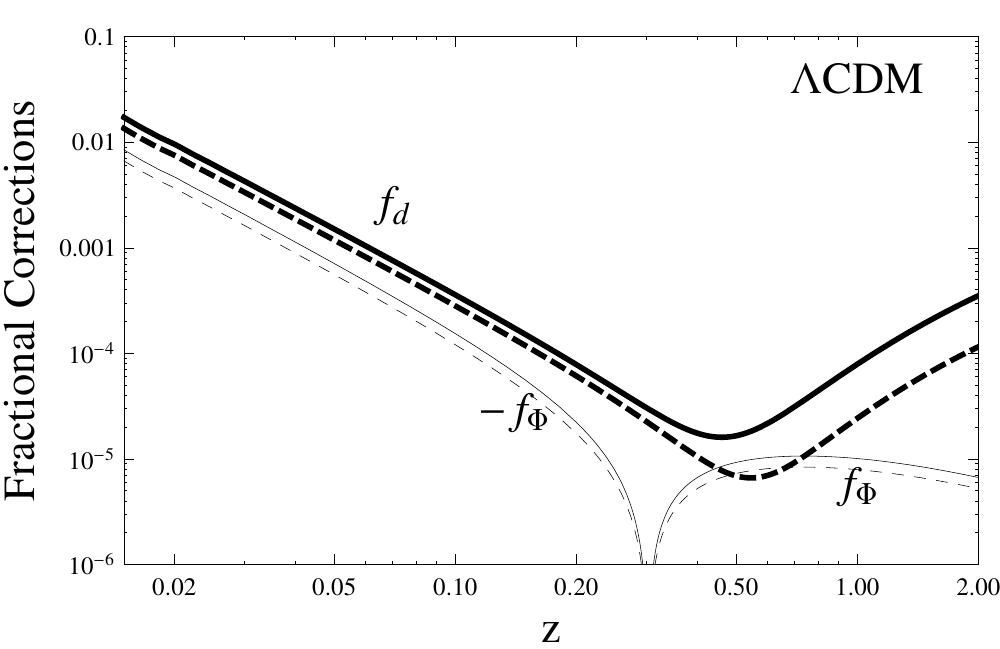}
\centering
\caption{The fractional correction to the flux $f_\Phi$ of Eq. (\ref{7a}) (thin curves) is plotted together with the fractional correction to the luminosity distance $f_d$  of Eq. (\ref{13}) (thick curves), for a $\La$CDM model with $\Om_{\La 0}=0.73$. We have used two different cutoff values: $k_{UV}=0.1 {\rm Mpc}^{-1}$ (dashed curves) and $k_{UV}=1 {\rm Mpc}^{-1}$ (solid  curves). The left panel shows the results obtained with a linear spectrum without baryon contributions. The right panel illustrates the effects of including baryons
(we have used, in particular,  $\Omega_{b0}=0.046$).}
\label{Fig3}
\end{figure}

The results of such a computation are illustrated in Fig. \ref{Fig3}, and a comparison with Fig. \ref{f1} clearly shows that $|f_\Phi|$ is smaller in the  $\Lambda$CDM case than in the CDM case, and further (slightly) depressed when we take into account the presence of a small  fraction of baryon matter (the curves presented in the right panel). In any case, the small values of $|f_\Phi|$ at relatively large $z$, for a realistic $\Lambda$CDM scenario,  lead us to conclude that the averaged flux is a particularly appropriate quantity for extracting from the observational data the ``true" cosmological parameters. As we will discuss now, the situation is somewhat different for other functions of $d_L$. 

%%%%%%%%%%%%%%%%%%%%%%%%%%%%%%

%%%%%%%%%%%%%%%%%%%%%%%%%%%%%%%%%%%%%%%%%%%%%%%%%%%%%%%%%%%%%%%%%%%%%%%%%%%%%%%%%%%%%%%%%%%%%%%%%%%%%%%

\subsection{Second-order corrections to other observables and dispersions}
\label{Sec6B}
%%%%%%%%%%%%%%%%%%%%%%%%%

Let us now consider other observables, beyond the flux, to see  how the impact of the inhomogeneities may change. We will treat, in particular, the two important examples introduced in Sect. \ref{Sec3}, namely the luminosity distance $d_L$ and the distance modulus $\mu$. The fractional corrections to their averages (see Eqs. (\ref{2}--\ref{14})) are qualitatively different from those of the averaged flux (represented by $f_\Phi$), because of the presence of  extra  contributions, unavoidable for any non-linear function of the flux and  proportional to the square of the first order fluctuation $(\Phi_1/\Phi_0)^2$. 

For a better understanding of such contributions let us start with the results obtained in \cite{BMNV}, concerning the second-order perturbative expansion of the luminosity distance, $d_L= d_{L}^{(0)}+
 d_{L}^{(1)}+ d_{L}^{(2)}$, and summarized in Sect. \ref{Sec2B} (see in particular Eqs .(\ref{215}) and (\ref{Ideltarel})). Using those results we obtain:
\be 
\frac{\Phi_1}{\Phi_0}=-2 \frac{d_{L}^{(1)}}{d_{L}^{(0)}}=
 - {\cal I}_1 + ({\rm t.d.})^{(1)} = 
2\left(-\Xi_s J+\frac{Q_s}{\Delta \eta}+\psi_s+J_2^{(1)}\right)\,.
\label{OBcorrection}
\ee
In order to determine the leading corrections we first notice that, by applying the  procedure of Sect. \ref{Sec5}, and computing the averages for the CDM case, we obtain, to leading order,
\be 
\overline{\left\langle \left(\frac{\Phi_1}{\Phi_0}\right)^2 \right\rangle}_{ L}=4\left\{\overline{\left\langle \left(J_2^{(1)}\right)^2 \right\rangle}+
\Xi_s^2 \left[\overline{\langle  ([\partial_r P]_s)^2  \rangle}+\overline{\langle  ([\partial_r P]_o)^2  \rangle}\right]\right\}\,.
\label{615}
\ee
Working in the context of a $\Lambda$CDM model, considering only these leading terms, and limiting ourselves to the linear regime, we find that the terms multiplying $\Xi_s^2 $ on the right hand side of the above equation can be calculated without approximations 
(as seen in the previous subsection).
 Also, we find that their contribution is controlled by the spectral factor $k^2 {\cal P}_\psi(k, \eta_o)$. The first term, on the contrary, is due to  to the so-called ``lensing effect", dominates at large $z$ (as already  shown in \cite{BGMNV1} for a CDM model), and has leading spectral contributions of the type $k^3 {\cal P}_\psi(k, \eta_o)$. For such term, however, the integrals over time cannot be factorized with respect to the $k$ integrals, and we must use the approximation already introduced in the previous subsection (namely, we have to replace in the integrands $\psi(\eta',\eta_o-\eta', \tilde{\theta}^a)$  with $\psi(\eta_s,\eta_o-\eta', \tilde{\theta}^a)$). The full result for the spectral coefficient can then be finally written as follows:
\be
\Ccal(\left(\Phi_1/\Phi_0\right)^2_L)\simeq \frac{4}{3}\Xi_s^2 \left(\ti f_s^2 + \ti f_o^2\right)k^2 +
\frac{4}{15} \left[\frac{g(\eta_s)}{g(\eta_o)}\right]^2 \Da \eta^3 k^3\,{\rm SinInt}(k \Da \eta) \,. 
\ee

Because of the new term (due to lensing) affecting the average of $\mu$ and $d_L$ -- but not of the flux $\Phi$ -- we may expect larger fractional corrections for these variables, as well as for other functions of $\Phi$, at higher redshifts. This is indeed confirmed by the plots presented Fig. \ref{Fig3} reporting the results of an explicit numerical integration and comparing, in particular,  the value of $f_d$ with the absolute value of $f_\Phi$. We obtain $|f_\Phi| \ll f_d$, at large values of $z$ where the lensing term dominates, both in the presence and in the absence of the baryon contribution to the total energy density. It should be stressed, however, that also the new $k^3$-enhanced contributions are free from IR and UV divergences, at least for the class of models we are considering.

\begin{figure}[t]
\centering
\includegraphics[width=8cm]{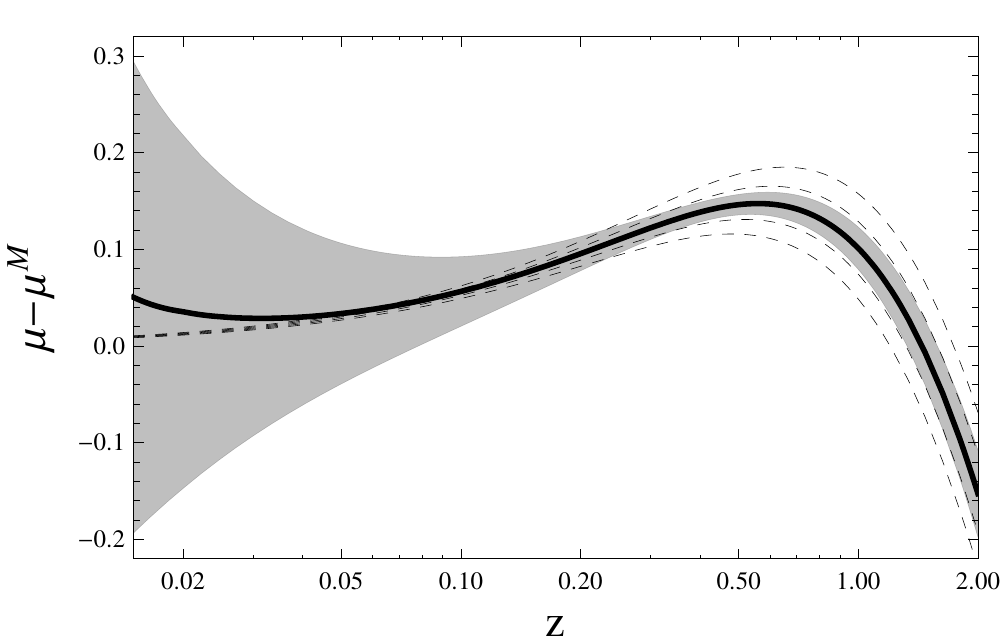}~~~~~~~~~
\includegraphics[width=8cm]{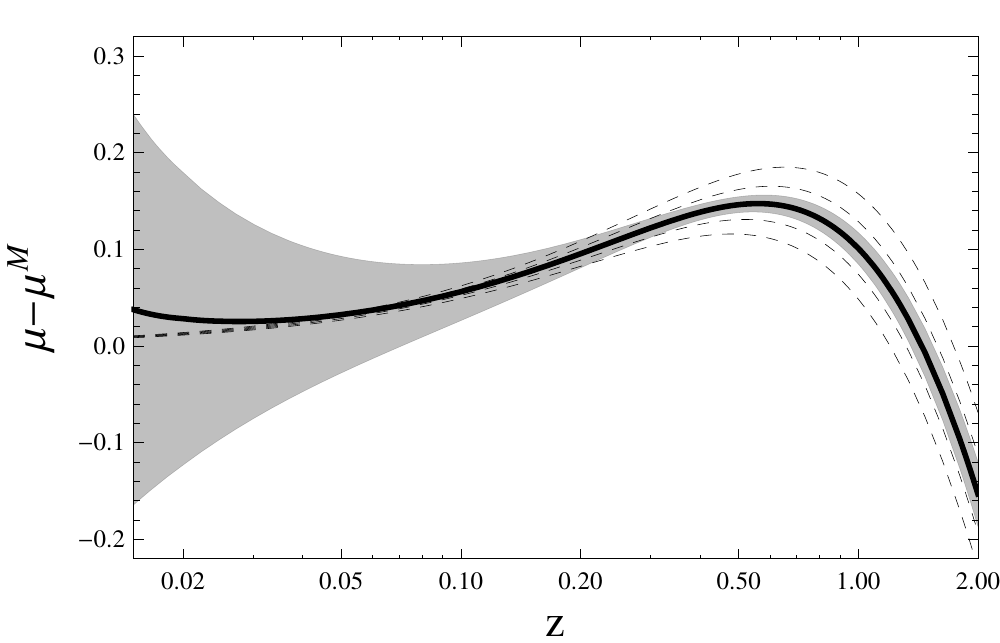}
\centering
\caption{The averaged  distance modulus 
$ \overline{\langle \mu \rangle} -\mu^M$ of Eq. (\ref{14}) (thick solid curve), and its dispersion of Eq. (\ref{15}) (shaded region) are computed for $\Om_{\La 0}=0.73$ and compared with the homogeneous value  for the unperturbed $\La$CDM models with, from bottom to top,  
$ \Om_{\La 0}= 0.69$, $0.71$, $0.73$, $0.75$, $0.77$ (dashed curves). We have used  $k_{UV}=1\,\rm{Mpc}^{-1}$. The left panel shows the results obtained with a linear spectrum without baryon contributions. The right panel illustrates the effects of including baryons, with $\Omega_{b0}=0.046$.}
\label{Fig4}
\end{figure}

Let us now discuss to what extent the enhanced corrections due to the square of the first-order flux fluctuation can affect the determination of the dark-energy parameters, if quantities other than the flux are used to fit the observational data. To this purpose we may consider the much used (average of the) distance modulus given in Eq. (\ref{14}), referring it,  as usual, to a homogeneous Milne model with $\mu^M= 5 \log_{10} [(2+z)z/(2H_0)]$.  
Considering Eqs. (\ref{14}) and (\ref{15}), where the averaged value of the distance modulus and its dispersion are given in function
of $f_\Phi$ and of $\overline{\left\langle \left({\Phi_1}/{\Phi_0}\right)^2 \right\rangle}$, we have investigated the magnitude of the effect in this case. 
In particular, in  Fig. \ref{Fig4} we have compared  the averaged value $ \overline{\langle \mu \rangle} -\mu^M$ with the corresponding expression for  homogeneous $\La$CDM models with different values of $\Om_{\La 0}$. We have also illustrated the expected dispersion around the averaged result, represented by the dispersion previously reported in Eq. (\ref{15}) (and already computed in \cite{BGMNV1} for the CDM case). 

We have found that the  given inhomogeneities, on the average, may affect the determination of $\Om_{\La 0}$ obtained from the measure of the distance modulus, at large $z$, only at the third decimal figure (at least if the spectral contributions are computed in the linear  regime). As we can see from Fig. \ref{Fig4}, the curves for $\overline{\langle \mu \rangle}$ and for the corresponding unperturbed value $\mu^{\rm FLRW}$ (with the same $\Om_{\La 0}$) practically coincide at large enough $z$. It should be stressed, also, that the dispersion on the distance modulus computed from Eq. (\ref{15}) reaches,  at large redshift,  a value which is
comparable with a change of about 2\%  in the dark energy parameter
$\Omega_{\Lambda 0}$ (cf.  \cite{BGMNV2}, considering however that baryonic effects have now been added
in the transfer function). We shall see in the next section that this
effect is enhanced by the use of  non-linear power spectra. 

%%%%%%%%%%%%%%%%%%%%%
%%%%%%%%%%%%%%%%%%%%%

%%%%%%%%%%%%%%%%%%%%%%%%%%%%%%%%%%%%%%%%%%%%%%%%%%%%%%%%%%%%%%%%%%%%%%%%%%%%%%%%%%%%%%%%%

\section{The $\Lambda$CDM model:  power spectrum in the non-linear regime}
\label{Sec7}
\setcounter{equation}{0}
%%%%%%%%%%%%%%%%%%%%%%%%%%%%%%%%%%%%%%%%%%%%%%%%%%%%%%%%%%%%%%%%%%%%%%%%%%%%%%

The linear spectra considered so far are sufficiently accurate only up to scales of order $0.1 h\, {\rm Mpc}^{-1}$. 
If we want to better study the effect of shorter-scale inhomogeneities on our light-cone averages we need to go beyond such linear approximation, taking into consideration the non-linear  evolution of the gravitational perturbations. This can be done by using the so-called ``HaloFit" models, which are known to reproduce quite accurately the results of cosmological $N$-body simulations. In particular, the HaloFit model of  \cite{Smith:2002dz} and its recent upgrade of  \cite{Takahashi} provide an accurate fitting formula for the power spectrum up to wavenumber $k \simeq 30 h \,{\rm Mpc}^{-1}$. 

In our previous paper \cite{BGMNV2} the analysis was limited to $k<1 \,{\rm Mpc}^{-1}$, but with the use of the non-linear power spectra we can now extend our analysis up to the maximum scale of the mentioned HaloFit models. The main difficulty with such an extension is that the time (i.e. $z$) dependence of the spectrum becomes more involved than in the linear case, since different scales no longer evolve independently. Hence, the need for introducing  approximations in performing the integrals becomes even more essential in this case. 

Let us start by recalling a few details of the HaloFit model. The fractional density variance per unit $\ln k$ is represented by the variable $\Delta^2(k)$,  defined by \cite{Smith:2002dz}:
\be
 \sigma^2 \equiv \overline{\delta(x)\delta(x)} = \int \frac{d^3k}{(2 \pi)^3} |\delta_{\vec{k}}|^2   = \int \Delta^2(k) ~ d\ln k ~,~
\ee
which implies
\be
\label{Deltadelta}
\Delta^2(k) = \frac{k^3}{2 \pi^2} |\delta_k|^2 \,.
\ee
where $\delta(x)= \da \r(x)/\r$ is the fractional density perturbation of the gravitational sources, and $\da_k$ is the associated Fourier component. 
On the other hand, the power spectrum of scalar perturbations, $\Pcal_\Psi(k,z)$, is related to $\Delta^2(k,z)$ by the Poisson equation, holding at both the linear ($L$) and non-linear ($NL$) level \cite{PeterUzan}:
\be
\label{PE}
\Pcal_\psi^{\rm{L,NL}}(k,z) = \frac{9}{4} \frac{\Om_{m0}^2 \Hcal_0^4}{k^4} (1+z)^2 \Delta_{\rm{L,NL}}^2(k,z) ~.
\ee
The linear part of the spectrum is used to introduce  the normalization equation, defining the non-linearity length scale  $k_\sigma^{-1}(z)$, as follows:
\be
\sigma^2(k_\sigma^{-1}) \equiv \int \Delta^2_{\rm{L}}(k,z) \, \exp(-(k/k_\sigma)^2)\;d\ln k \equiv 1. 
\ee
It is then obvious that the scale $k_{\sigma}$ is redshift-dependent. Since the non-linear power spectra  obtained from the models  \cite{Smith:2002dz} and \cite{Takahashi} are using such a scale, they will be characterized by an implicit  $z$-dependence which is more complicated than the one following from the usual growth factor of Eq. (\ref{gformula}), and which cannot be factorized. 

Besides $k_{\sigma}$, the linear spectrum also determines two additional  parameters, important for the construction of the HaloFit model:
the effective spectral index  $n_{\rm eff}$ and the parameter $C$, controlling  the curvature of the spectral index at the scale $k_{\sigma}$. They are defined by:
\be
3+n_{\rm eff} \equiv -\left.{d \ln \sigma^2(R) 
\over d \ln R}\right|_{R=k_\sigma^{-1}} ~~,~~
C \equiv - \left.{d^2 \ln \sigma^2(R) \over d \ln R^2}
\right|_{R=k_\sigma^{-1}} ~~.
\ee
Once the values of $k_\sigma$, $n_{\rm eff}$ and $C$ are determined, they can be inserted into a given HaloFit model \cite{Smith:2002dz,Takahashi}  to produce the non-linear power spectrum $\Delta_{\rm{NL}}^2(k,z)$. One finally goes back to $\Pcal_\psi^{\rm{NL}}(k,z)$ using again the Poisson equation (\ref{PE}).

\begin{figure}[t]
\centering
\begin{tikzpicture}
\node (label) at (0,0){
        \includegraphics[width=14cm]{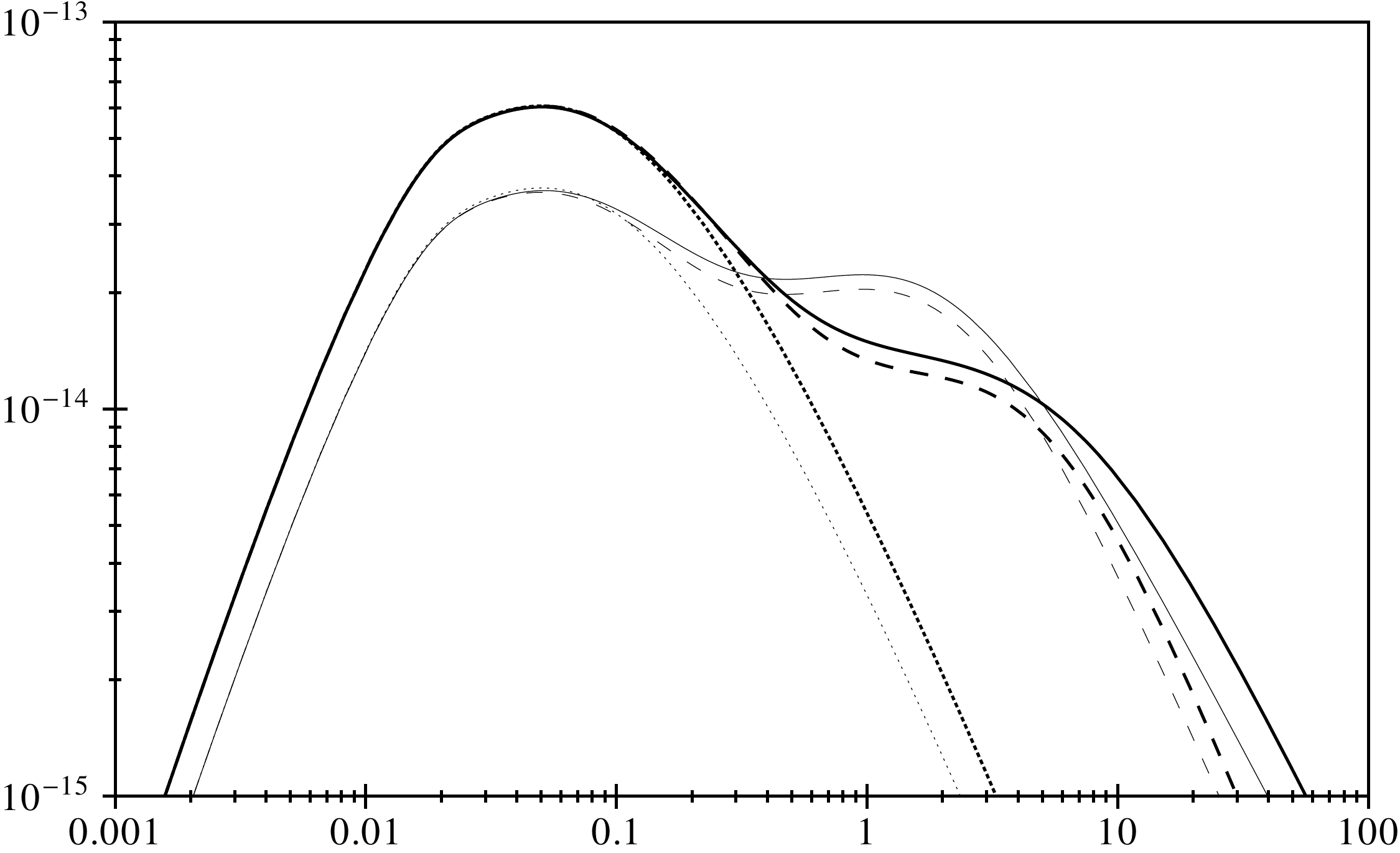}
      };

\node[anchor=west] at (-5.5,1) (description) {\Large $\textbf{z=1.5}$};
\node[anchor=west] at (-4,-2) (description) {\Large $\textbf{z=0}$};

\node[anchor=west] at (3.8,0.8) (description) {\Large $\textbf{k}^2 ~ \Pcal_{\Psi}^{\NL}(\textbf{k})$};
\node[anchor=west] at (-0.5,-2.5) (description) {\Large $\textbf{k}^2 ~ \Pcal_{\Psi}^{\L}(\textbf{k})$};

\node[anchor=west,rotate=90] at (-7.5,-2.5) (description) {\Large $\textbf{k}^2 ~ \Pcal_{\Psi}^{\L,\NL}(\textbf{k}) ~~[h^2 ~ {\rm Mpc}^{-2} ]$};

\node[anchor=west] at (-1,-4.7) (description) {\Large $\textbf{k} ~[h ~{\rm Mpc}^{-1}]$};
\end{tikzpicture}
\caption{The linear spectrum $\Pcal_\psi^{\rm{L}}$ (dotted curves) and the non-linear spectrum $\Pcal_\psi^{\rm{NL}}$ for the HaloFit model of \cite{Smith:2002dz} (dashed curves) and of \cite{Takahashi} (solid curves). In all three cases the spectrum is multiplied by $k^2$ (for graphical convenience) and is given for $z=0$ (thin curves) and $z=1.5$ (thick curves). We have included  baryons with $\Om_{b0}=0.046$.}
\label{Fig5}
\end{figure}
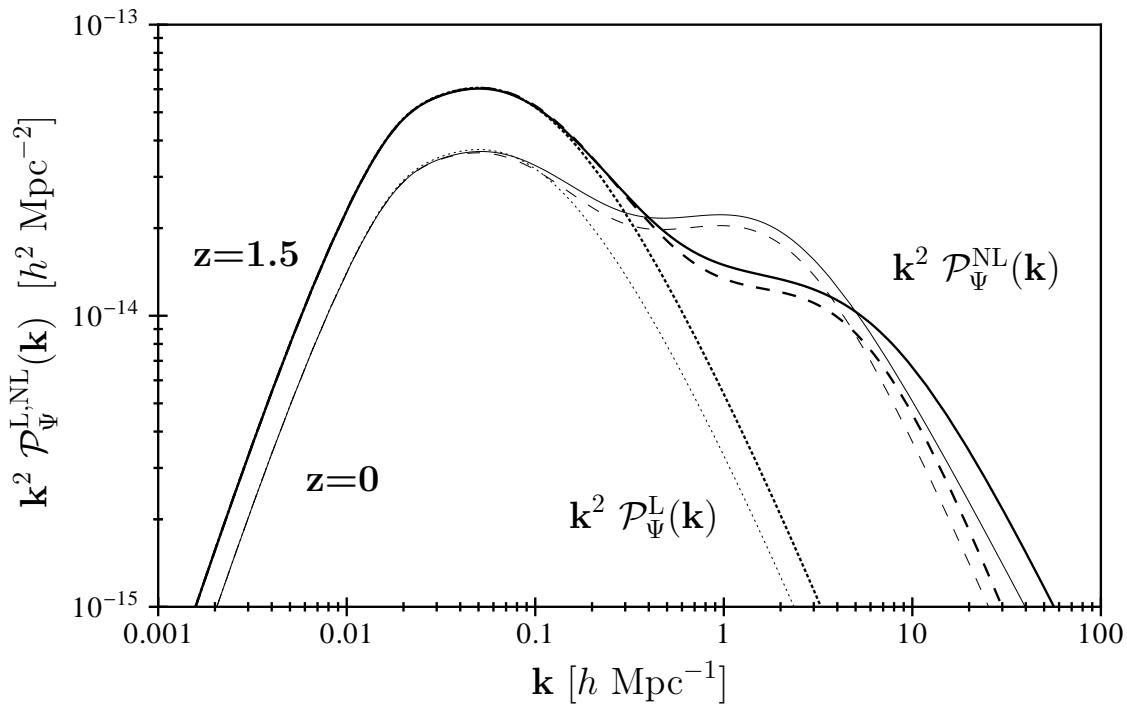

The non-linear spectra obtained in this way, and based on the previous linear spectrum with baryon contributions, are illustrated in Fig. \ref{Fig5}  for the two HaloFit models \cite{Smith:2002dz} and \cite{Takahashi}. The  results are compared with the spectrum of the linear regime, for different values of the redshift ($z=0$ and $z=1.5$). We can see that the spectra  intersect each other in the non-linear regime ($k\gaq 0.1 h\, {\rm Mpc}^{-1}$), 
as a result of the intricate redshift dependence. We  can also observe that  the two HaloFit models lead to the same result at low values of $k$ (including baryons, but without BAO).

%%%%%%%%%%%%%%%%%%%%%%%%%%%%%%%%%%%%%%%%%%%%%%%%%%%%%%%%%%%%%%%%%%%%%%
\subsection{Numerical results and comparison with the linear regime}
\label{Sec7A}
%%%%%%%%%%%%%%%%%%%%%%%%%%%%%%%%%%%%%%%%%%%%%%%%%%%%%%%%%%%%%%%%%%%%%%%%%%%%%

In order to evaluate our  integrals using the non-linear power spectra we need  to face the additional (already mentioned)  problem due to the fact  that  the non-linear spectra -- unlike the linear ones -- cannot be factorized  as a function of $k$ times a function of $z$. In that case the full two-dimensional integration is highly non-trivial, and we have thus exploited a further approximation. In the presence of time-integrals of the mode functions (like those appearing, for instance, in the leading terms of $\overline{\langle \Ical_{1,1} \rangle}$), we have parametrized the non-linear power spectrum in a factorized form as follows:
\beq
\Pcal_\Psi^{\rm{NL}}(k, z) = \frac{g^2(z)}{g^2(z^*)} \Pcal_\Psi^{\rm{NL}}(k, z^*),
\label{76}
\eeq
and we have chosen $z^*=z_s/2$ to try to minimize the error. 

Let us briefly comment on the validity of the approximation we have introduced.
Our  analysis being focussed on the redshift range $0.015 \leq z \leq 2$,  the above approximation is most inadequate only when we consider $z_s=2$ (i.e. $z^*=1$), and our mode functions are evaluated inside the integrals at the values of $z$ most distant from $z^*$ (namely, $z=0.015$ and $z=2$). In that case we are lead to underestimate the spectrum by about a $40\%$ factor for $z=0.015$, and to overestimate it by about a $80\%$ factor for $z=2$. However, this only occurs in two  narrow bands of $k$ centered around the values $k=1 h\,{\rm Mpc}^{-1}$  and $k=2 h \,{\rm Mpc}^{-1}$,  
while, outside these bands, our approximation is good. Since these errors are limited to only  a part of the region of integration, both in $z$ and in $k$, we can estimate an overall accuracy at the $10\%$ level, at least,  for the results given by the adopted approximation. We have also checked the sensitivity of the numerical results to changes in $z^*$, such as $z^*=z_s$, and checked that our final results are only weakly dependent (typically at the $1\%$ level) on the  choice of $z^*$.

\begin{figure}[ht!]
\centering
\includegraphics[width=13cm]{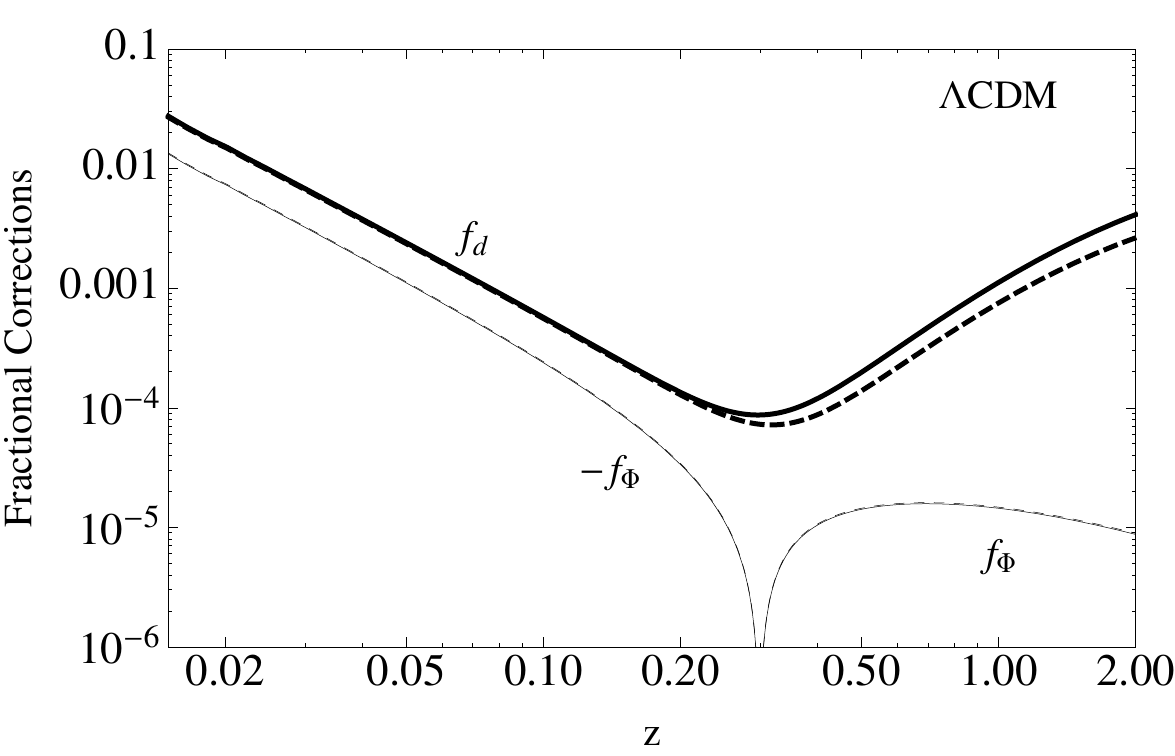}
\centering
\caption{The fractional correction to the flux ($f_\Phi$, thin curves) and to the luminosity distance ($f_d$, thick curves), for a perturbed $\La$CDM model with $\Omega_{\Lambda 0}=0.73$. Unlike in Fig. \ref{Fig3}, we have taken into account the non-linear contributions to the power spectrum given by the HaloFit model of \cite{Takahashi} (including baryons), and we have used the following cutoff values: $k_{UV}=10 h \,{\rm Mpc}^{-1}$ (dashed curves) and $k_{UV}=30 h \,{\rm Mpc}^{-1}$ (solid  curves).}
\label{Fig6}
\end{figure}

\begin{figure}[ht!]
\centering
\includegraphics[width=8.2cm]{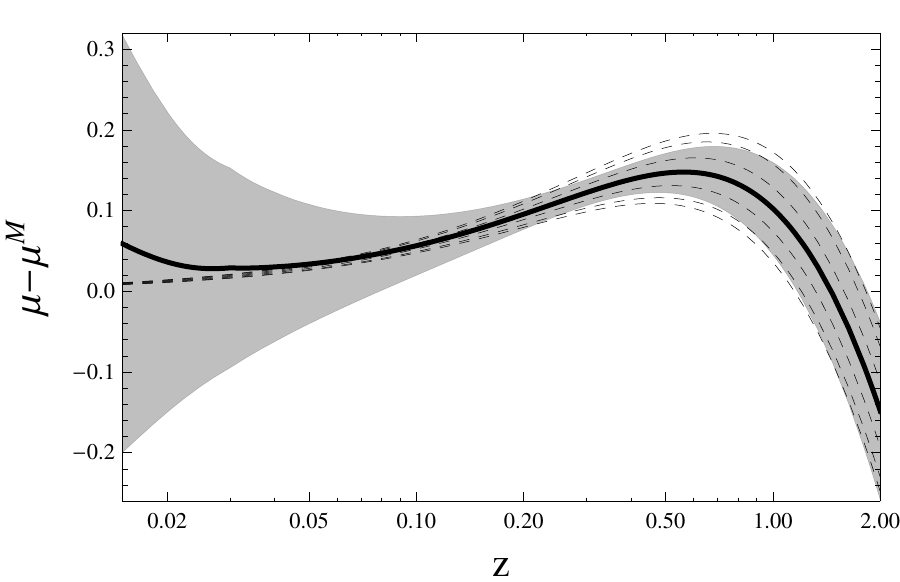}~~~~~~~~~
\includegraphics[width=8.2cm]{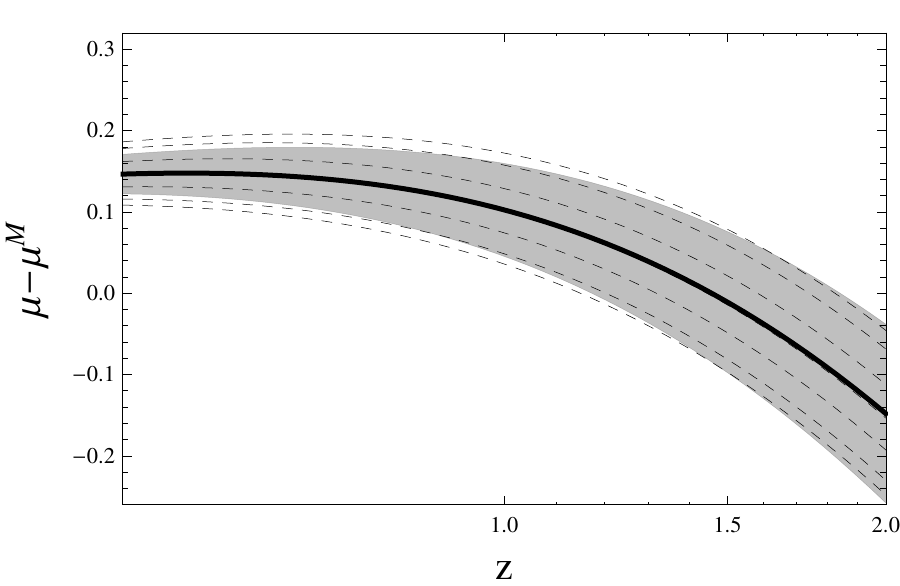}
\centering
\caption{The averaged  distance modulus $ \overline{\langle \mu \rangle} -\mu^M$ of Eq. (\ref{14}) (thick solid curve), and its dispersion of Eq. (\ref{15}) (shaded region), for a perturbed $\La$CDM model with $\Omega_{\Lambda 0}=0.73$. Unlike Fig. \ref{Fig4}, we have taken into account the non-linear contributions to the power spectrum given by the HaloFit model of \cite{Takahashi} (including baryons), and used the cut-off  $k_{UV}=30 h \,{\rm Mpc}^{-1}$. The averaged results are compared with the homogeneous values of $\mu$ predicted by  unperturbed $\La$CDM models with  (from bottom to top) $\Omega_{\Lambda 0}= 0.68$, $0.69$ $0.71$, $0.73$, $0.75$, $0.77$, $0.78$ (dashed curves). The right panel simply provides a zoom of the same curves, plotted in the smaller redshift range $0.5 \leq z \leq 2$.}
\label{Fig7}
\end{figure}

Once we have established the range of validity of the parametrization (\ref{76}),    we  proceed in evaluating  the $z$ (or $\eta$) integrals as we did in Sect. \ref{Sec6}. The final results of this procedure are illustrated by the curves plotted in Figs. \ref{Fig6} and \ref{Fig7}, computed with the non-linear power spectrum following from  the HaloFit model of \cite{Takahashi} and including baryon contributions\footnote{The two HaloFit models quoted above give similar  results, and we have chosen to present here, for simplicity, only those obtained with one of them.}.
As illustrated for instance in Fig. \ref{Fig6}, the fractional correction to $d_L$ turns out to be of order of a few parts in $10^{-3}$ around $z=2$, and smaller in the rest of the intermediate redshift range relevant for cosmic acceleration. 
On the other hand, in the same redshift range, the fractional correction to the flux is about two orders of magnitude smaller.

Comparing with the results obtained with the linear spectrum  (see Figs. \ref{Fig3}, \ref{Fig4}), we can see that taking into account the non-linearity distortions (and using higher cut-off values) enhances the backreaction effects on the considered functions of the luminosity distance, but not enough to reach a (currently)  observable level. On the other hand, the dispersion, already large in the linear case, is further  enhanced when non-linearities are included. In particular, the dispersion on $\mu$ due to inhomogeneities is of order $10\%$ around $z=2$, which implies that the predictions of homogeneous models with 
$\Omega_{\Lambda 0}$ ranging from $0.68$ to $0.78$ lie inside one standard deviation with respect to the averaged predictions of a perturbed (inhomogeneous) model with $\Omega_{\Lambda 0}=0.73$.

%%%%%%%%%%%%%%%%%%%%%%%%%%%%%
%%%%%%%%%%%%%%%%%%%%%%%%%%%%%

%%%%%%%%%%%%%%%%%%%%%%%%%%%%%
%%%%%%%%%%%%%%%%%%%%%%%%%%%%%

%%%%%%%%%%%%%%%%%%%%%%%%%%%%%
%%%%%%%%%%%%%%%%%%%%%%%%%%%%%

%%%%%%%%%%%%%%%%%%%%%%%%%%%%%%%%%%%%%%%%%%%%%%%%%%%%%
\subsection{Comparing theory and observations via  the intrinsic dispersion of the data}
 \label{Sec7B}
%%%%%%%%%%%%%%%%%%%%%%%%%%%%%%%%%%%%%%%%%%%%%%%%%%%%%

\begin{figure}[t]
\centering
\includegraphics[width=12cm]{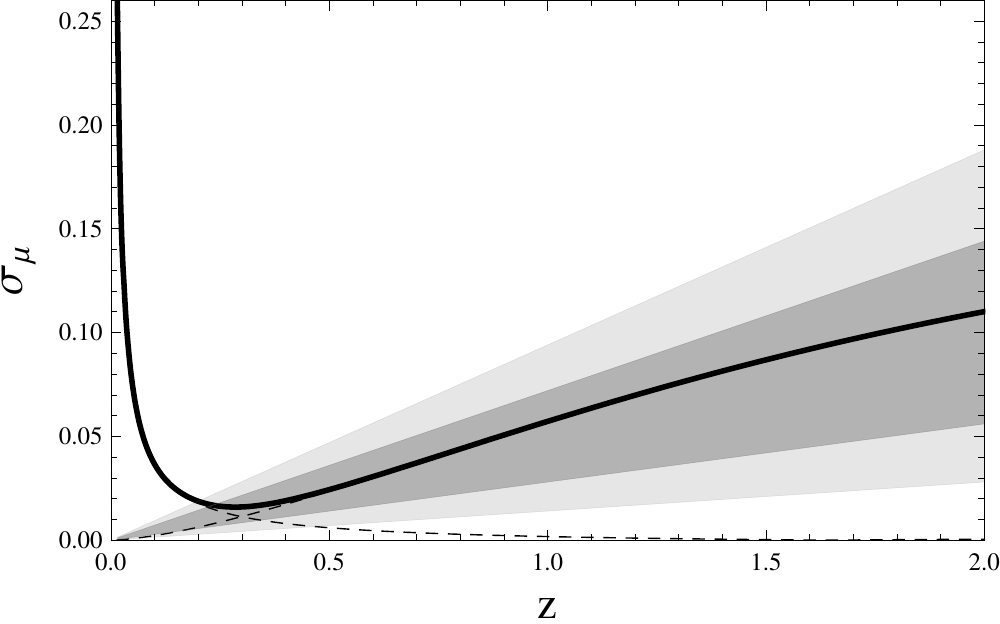}
\centering
\caption{The $z$-dependence of the total dispersion $\sg_\mu$ is illustrated by the thick solid curve, and it is separated into its ``Doppler" part (dashed curve dominant at low $z$)  and ``lensing" part (dashed curve dominant at large $z$). The slope of the dispersion in the lensing-dominated regime is compared with the experimental estimates of  Kronborg \etal \cite{Kronborg} (dark shaded area), and  of J\"onsson \etal \cite{Jonsson} (light shaded area).}
\label{Fig8}
\end{figure}

Let us now consider in more detail our prediction for the dispersion $\sigma_{\mu}$ induced by the presence of the inflationary perturbation background, and compare it with the intrinsic dispersion of the distance modulus that can be inferred from SNe Ia data. Our results for  the dispersion are already implicitly contained in Fig. \ref{Fig7}  but, for the sake of clarity,  we have  separately plotted our value of $\sg_\mu$  in Fig. \ref{Fig8}, where the thick solid curve represents the value of $\sg_\mu$ obtained from Eq. (\ref{15}), and plotted as a function of $z$.  We can see from the figure that $\sigma_{\mu}$ has a characteristic $z$-dependence, with a minimal value of about $0.016$ reached around $z = 0.285$. Also,  the total dispersion of Eq. (\ref{15}) nicely interpolates between the leading Doppler contribution obtained from the two last terms of Eq. (\ref{615}) (represented  by the dashed curve approaching zero at large $z$), and the leading lensing contribution obtained from  the first term of Eq. (\ref{615}) (represented  by the dashed curve approaching zero at small $z$). 

The total variance $\sigma_{\mu}^{\rm obs}$ associated with the observational data,  on the other hand, can be decomposed in general  as follows (see e.g. \cite{March:2011xa,SNLS3,VN}): 
\beq
(\sigma_{\mu}^{\rm obs})^2 =  (\sigma_{\mu}^{\rm fit})^2 + (\sigma_{\mu}^{z})^2  + (\sigma_{\mu}^{\rm int})^2 ~~.
\eeq
Here $ \sigma_{\mu}^{\rm fit}$ is  the statistical uncertainty due, for instance, to the method adopted for fitting the light curve (e.g. the so-called SALT-II method
\cite{Guy}), but also to the uncertainty in the modeling of the supernova process. The term $\sigma_{\mu}^{z}$ represents instead the uncertainty in redshift due to the peculiar velocity of the supernova  as well as to the precision of spectroscopic measurements. 
Finally, $\sigma_{\mu}^{\rm int}$ is an unknown phenomenological quantity, needed to account for the remaining dispersion of the data with respect to the chosen homogeneous model. This part of the dispersion can be subsequently redefined whenever we are able to estimate some of the possible contributions it contains. The contribution we are mainly interested in here is the one originating from the lensing effect, which is dominant at large redshift. We can thus write, at large $z$:
\beq
(\sigma_{\mu}^{\rm int})^2 = (\widehat{\sigma_{\mu}^{\rm int}})^2 + (\sigma_{\mu}^{\rm lens})^2  ~~,
\eeq
where  $\widehat{\sigma_{\mu}^{\rm int}}$ is the remaining source of intrinsic dispersion.

Given the typical precision of current data \cite{Kowalski,Guy}, a reasonable fit of the Hubble diagram does not seem to require a strong $z$-dependence of the parameter $\sigma_{\mu}^{\rm int}$: for instance, a nearby sample gives $\sigma^{\rm int} = 0.15 \pm 0.02$, to be compared with the value $\sigma^{\rm int} = 0.12 \pm 0.02$ obtained for distant supernovae.  
On the other hand, as illustrated in Fig. \ref{Fig8}, the results of our computations at $z \gaq 0.3$ is very well captured by a linear behaviour which can be roughly fitted by $\sigma_{\mu}^{\rm lens}(z) = 0.056 z$. We should also note that this contribution stays below $0.12$ up to $z \sim 2$, which makes it perfectly compatible with observations so far performed.

It is remarkable that the (above mentioned) simple linear fit of our curve for $\sg_\mu(z)$ at $z \gaq 0.3$ turns out to be very close to the experimental estimate reported in \cite{Kronborg}, namely $\sigma_{\mu}^{\rm lens}=(0.05 \pm 0.022) z$. Also, such a fit is well compatible with the results of \cite{Jonsson}, namely $\sigma_{\mu}^{\rm lens}=(0.055^{+0.039}_{-0.041})z$. These two observational estimates of the lensing dispersion, with the relative error bands, are illustrated by the shaded areas of Fig. \ref{Fig8}. Our result is also in relatively good agreement with the simulations carried out in \cite{Holz} which predicted an effect of $0.088 z$. Other authors (\cite{Williams, Menard, Karpenka}), however, have found no indication of this $z$-dependence of $\sg_\mu$, and it is true that such a signal waits for a better observational evidence. 

It is likely that future improvements in the accuracy of SNe data will detect (or disprove) this effect at a higher confidence level. In this respect, our results  for $\sg_\mu(z)$ stand out as a challenging prediction\footnote{We remark, incidentally, that our prediction for the Doppler-related dispersion at small $z$ is also consistent with previous findings \cite{HG} on the so-called  Poissonian  peculiar-velocity contribution to $\sigma_{\mu}$. We have checked that our result for the Doppler contribution (see Fig. \ref{Fig8}) is well fitted by an inverse power law: $\sg_\mu(z)  \sim 0.00323 z^{-1} $ in very good agreement, shape-wise, with the corresponding result in  \cite{HG}. The fact that our prediction is about  a factor 1.7 larger than the one of \cite{HG}  is due, presumably, to the use of somewhat different power spectra (linear or non-linear, with or without baryons).}, which, we believe, could represent a further significative test of the concordance model.

%%%%%%%%%%%%%%%%%%%%%%%%%%%%%%%&&&
%%%%%%%%%%%%%%%%%%%%%%%%%%%%%%%%%%%%%%%%%%%%%%%%%%%%%%%%%%%%%%%%%%%%%%

%%%%%%%%%%%%%%%%%%%%%%%%%%%%%%%%%%%%%%%%%%%%
\section{Conclusions}
\label{Sec8}
\setcounter{equation}{0}
%%%%%%%%%%%%%%%%%%%%%%%%%%%%%%%%%%%%%%%%%%%%%%%%%%%%%%%%%%%%%%%%%%%%%%%%%%%%%%

Starting from the result of a previous paper \cite{BMNV},  where the luminosity-redshift relation has been computed to second order in the Poisson gauge, we have proceeded here to the evaluation of the effects of a realistic stochastic background of perturbations on the determination of dark energy parameters.
The basic tool we have used is the gauge-invariant light-cone averaging procedure proposed in \cite{GMNV}, applied to different functions of the luminosity distance $d_L$  averaged over a constant-redshift surface lying on our past light-cone. 

As already explained in \cite{BGMNV1}, different functions of $d_L$ differ by  the  sensitivity of their light-cone averages to fluctuations. Remarkably, a directly observable variable like the luminosity flux, $\Phi \sim d_L^{-2}$ turns out to be the least sensitive to perturbations. Its averaged expression is also the simplest and, fortunately enough, other averages are readily computed once the one of the flux is given. Similarly, calculation of the dispersion is straightforward.

As far as modeling inhomogeneities goes,  we have used a concordance $\Lambda$CDM model with an arbitrary 
$\Omega_{\Lambda 0}$, upon which we have added a realistic spectrum of stochastic scalar perturbations up to second order (and we have also explained why, to this order, we have no contribution from vector and tensor perturbations). The perturbations have been taken as those originating from a quasi-scale-invariant  primordial spectrum with a realistic transfer function (including baryons and the relative Silk-damping) after it undergoes a non-linear evolution  according to the so-called HaloFit model for structure formation.
This model appears to agree well with numerical $N$-body simulations as well as with large-scale structure data.

Our main conclusions (already succinctly presented in \cite{BGMNV2} for the case of a perturbation spectrum computed in the linear regime) are that  the effect of perturbations on the averaged flux are extremely small, typically of order $10^{-5}$ at $z \sim O(1)$. Thus the average flux stands out as an extremely  safe observable for determining dark-energy parameters using the simplest FLRW geometry.  Such observable is also practically insensitive (see Fig. 
\ref{Fig6}) to the short-distance behaviour of the power spectrum. 
Other variables (like $d_L$ and the commonly used distance 
modulus $\mu$) receive corrections that are typically from two to 
three orders of magnitude larger (at large values of the redshift $z$), 
but still small-enough for allowing dark-energy measurements at the percent 
level without invoking theoretical corrections. On the other hand, they are more sensitive to the chosen value of the UV cutoff, 
and to the corrections arising from the power spectrum in the non-linear regime. The enhanced bias in these flux-related variables is simply due to the scatter in the flux at fixed $z$ (compare  Eqs. (\ref{13}) and (\ref{14}) with (\ref{16})). It is an effect that can (and needs to) be taken into account in the analyses of SNe data. The absence of this enhanced bias for the flux itself  confirms it as the best variable for all observational purposes. 

We find, however, that the predicted intrinsic dispersion (or scatter) of the data due to just stochastic inhomogeneities --and not to other well-known sources of dispersion-- is considerably larger than their effect on averages. They imply that data should fluctuate in a band which, at large redshifts, covers the FLRW luminosity curves corresponding to a spread in $\Omega_{\Lambda 0}$ of nearly $10 \%$ (see Fig. \ref{Fig7}). 
For limited statistics this irreducible dispersion will limit the precision with which dark-energy parameters can be extracted from the data. Particularly interesting, at large redshift, is the scatter due to lensing. Such an effect has been observed and a linear  phenomenological fit to $\sigma_{\mu}$ has been proposed  \cite{Kronborg, Jonsson} with a slope $d \sigma_{\mu}/{dz} \sim 0.05$ but with large errors. 
Our theoretical prediction is well described (for the considered range of $z$) by the linear behaviour $ \sigma_{\mu}(z) \sim 0.056z$ which is not only consistent with the above phenomenological fits  but also  provides an interesting test of the concordance model if and when a more precise determination of $\sigma_{\mu}^{\rm lens}$ will become available. 
Also at small redshifts our (Doppler-induced) scatter, obeying an approximate inverse power law  
$\sg_\mu(z)  \sim 0.00323 z^{-1} $, looks compatible with observations and with previous theoretical estimates \cite{HG}. As a result of both effects we find the intrinsic dispersion  $ \sigma_{\mu}(z)$ to have a minimum of about  $0.016$ at $z \sim 0.285$.

In any case, our conclusion is that, when averaging is  applied to a physical observable within a well defined gauge invariant formalism, not even a small fraction of cosmological constant can be simulated by a stochastic (i.e. statistically homogeneous and isotropic) background of inhomogeneities. The situation is obviously different if one is willing to consider  a deviation from the ``almost scale-invariant" primordial spectrum, or a deterministic inhomogeneous and/or anisotropic cosmological model where we move along a very special geodesic, or one is ready to depart from the General Relativity framework.

An interesting property of our averaging procedure is that, unlike other (more formal) definitions \cite{CU,Kolb}, it  leads to results which, for any realistic inhomogeneity power spectrum, are free from IR as well as UV divergences. The former property is very likely related to the gauge invariance of our procedure which, by definition,  is unaffected by gauge artifacts due to super-horizon scales. 
Insensitivity to the UV regime is consistent with the intuition that very short-scale inhomogeneities should average out when  considering large-scale physical observables (it may fail, instead, for artificially-defined spatial averages). Yet, some sensitivity to the actual UV cutoff remains in quantities that are controlled by a high (e.g. 3rd) moment of the power spectrum.
Although, strictly speaking, we have only checked these nice properties up to second order, we strongly believe that they persist at higher orders as well.
Actually, the use of our particularly suitable geodesic light-cone gauge  \cite{GMNV} for performing light-cone averages  may allow for a non-perturbative treatment of the  backreaction problem.

%%%%%%%%%%%%%%%%%%%%%%%%%%%%%%%%%%

\section*{ACKNOWLEDGMENTS}

IBD would like to thank Eric Switzer and Pascal Vaudrevange for fruitful discussions.
GM would like to thank R. Durrer, E. Di Dio and V. Marra for useful discussions.
FN wishes to thank J. Guy and D. Hardin for interesting discussions about dispersion in the SNe Ia data.
FN and GV would like to thank R. Scoccimarro for  interesting discussions about non-linear spectra and the HaloFit model. 

IBD would like to acknowledge the hospitality of the Hebrew University where part of this work was carried out. 
GM has enjoyed the hospitality of the Department of Mathematics at Rhodes University and of the Astrophysics, Cosmology and Gravity Centre of the University of Cape Town during the completion of this work. GV and MG acknowledge the hospitality of the University of Geneva during the last stages of this work.

The research of IBD is supported by the German Science Foundation (DFG) within the Collaborative Research Center 676 "Particles, Strings and the Early Universe".
GM is supported by the Marie Curie IEF, Project NeBRiC - ``Non-linear effects and backreaction in classical and quantum cosmology".

%%%%%%%%%%%%%%%%%%%%%%%%%%%%%%%%%%%

\begin{appendix}

\renewcommand{\theequation}{A.\arabic{equation}}
\setcounter{equation}{0}

\section*{Appendix A. Second-order vector and tensor perturbations}
\label{AppA}

We have already stressed in Sect. \ref{Sec2B} that vector and tensor perturbations  automatically appear, at second order, sourced by the squared first-order perturbation terms. Hence, vector and tensor perturbations  must be included in a consistent second-order computation of the luminosity distance, even if their contributions is negligible at first order (as expected, in particular, for a background of super-horizon perturbations generated by a phase of slow-roll inflation). 

Working in the Poisson gauge, and moving to spherical coordinates $x^i=(r, \theta, \phi)$, we can rewrite the relevant part of the PG metric (\ref{PGmetricstandard}) as follows:
\beq
ds_{PG}^2 = a^2 \left[ -d\eta^2 + 2 v_i d\eta d x^i \right]+ 
a^2\left[(\gamma_0)_{ij} + \chi_{ij}\right] d x^i d x^j \, ,
\eeq
so that
\beq
g_{PG}^{\mu\nu}(\eta,r,\theta^a) = a^{-2}
\left(
\begin{array}{cc}
-1 & v^i \\
v^j & \gamma_0^{ij} - \chi^{ij} \\
\end{array}
\right) ~,
\eeq
where $\ga_0^{ij} = {\rm diag}(1, r^{-2}, r^{-2} \sin^{-2} \theta)$, and where we have called $v^i$ and $\chi^{ij}$ the vector and tensor perturbations written in spherical polar coordinates. They satisfy the conditions $\nabla_i v^i=0= \nabla_i \chi^{ij}$ and $\ga_0^{ij} \chi_{ij}=0$, where $\nabla_i$ is the covariant gradient of three-dimensional Euclidean space in spherical coordinates.

Following the same procedure as in the scalar case we can now evaluate the vector and tensor contributions to the coordinate transformation connecting Poisson and GLC gauge, and then express the perturbed GLC metric, up to second order, including the vector and tensor variables $v^i$ and $\chi^{ij}$. Such a detailed computation has already been performed, and its results presented in \cite{BMNV}. For the purpose of this paper it will be enough to recall here the vector and tensor contributions to the coordinate transformation between $\theta$ and $\ti \theta$:
\be
\tilde{\theta}^a = \tilde{\theta}^{a (0)}+\tilde {\theta}^{a (2)}=\theta^a + \frac{1}{2} \int_{\eta_+}^{\eta-} dx \left( \hat{v}^a(\eta_+,x,\theta^a) - \hat{\chi}^{ra}(\eta_+,x,\theta^a) + \hat{\gamma}^{ab}_0(\eta_+,x,\theta^a) \int_{\eta_+}^x dy~ \partial_b \hat{\alpha}^r (\eta_+,y,\theta^a) \right) ~,
\label{A1}
\ee
and to the $2 \times 2$ matrix $\ga^{ab}$ appearing in the GLC metric:
\bea
&&
\!\!\!\!\!\!\!\!\!\!\!\!\!\!
a(\eta)^2 \gamma^{ab} = \gamma_0^{ab} - \chi^{ab} +
\nonumber \\ 
& &+ \left[ \frac{\gamma_0^{ac}}{2} \int_{\eta_+}^{\eta_-} dx~ \partial_c \left( \hat{v}^b(\eta_+,x,\theta^a) - \hat{\chi}^{rb}(\eta_+,x,\theta^a) + \hat{\gamma}_0^{bd} (\eta_+,x,\theta^a) \int_{\eta_+}^{x} dy~ \partial_d \hat{\alpha}^r (\eta_+,y,\theta^a) \right) + (a \leftrightarrow b) \right] ,
\label{A2}
\eea
where 
$
\alpha^r \equiv ({v^r}/{2}) - ({\chi^{rr}}/{4})$.
Both results are needed, in fact, for the computation of the averaged flux (\ref{dLminus2}).

If we take into account $v^i$ and $\chi^{ij}$, and compute $d_L$ according to Eqs. (\ref{dLdA}) and (\ref{dAGLC}), we find that the right-hand side of Eq. (\ref{215}) has to be modified by the addition of a new term, $\da^{(2)}_{V,T} (z_s, \ti \theta^a)$, representing the effect of the vector and tensor part of the perturbed geometry (see \cite{BMNV} for its explicit expression). No modification is induced, however, on the corresponding equation for $I_\phi(z_s)$ controlling the light-cone average of $d_L^{-2}$, so that Eq. (\ref{Iphi}) holds even in the presence of vector and tensor perturbations. 

The sought average, in fact, is proportional to the proper area of the deformed two-sphere $\Sg(w_o, z_s)$, and is given by $I_\phi \sim \int d^2 \ti\theta \sqrt{\ga}$ (see Eq. (\ref{dLminus2})). Considering the vector and tensor contributions to $\ga^{-1}= \det \ga^{ab}$ (obtained from Eq. (\ref{A2})), and computing from Eq. (\ref{A1}) the 
Jacobian determinant $| \pa \ti \theta/\pa \theta |$, we can  express $I_\phi$ as an angular integral over the two-sphere with unperturbed measure $d^2 \Om= \sin \theta d \theta d \phi$. In that case many terms cancel among each other, and we end up with the result:
\beq
I_\phi(w_o,z_s)-1 = {1\over 4 \pi} \int_0^\pi  \sin \theta d \theta \int_0^{2\pi} d\phi\, f(\eta, r, \theta, \phi),
\label{A3}
\eeq
where the integrand $f(\eta, r, \theta, \phi)$ is a simple expression  proportional to the components of the vector and tensor perturbations. 

The above angular integrals are all identically vanishing, as we can check by expanding the perturbations in Fourier modes $v^i_k$ and $\chi^{ij}_k$. For each mode 
$\vec k$ we can choose, without loss of generality, the $x^3$ axis of our coordinate system aligned along the direction of $\vec k$. Considering, for instance, tensor perturbations,  we can then write the most general perturbed line-element, in Cartesian coordinates (omitting, for simplicity, the Fourier index), as follows:
\beq
h_{ij} dx^i dx^j = h_{+}(\eta, x^3) (dx^1dx^1 - dx^2dx^2 ) +  2 h_{\times}(\eta,x^3) dx^1dx^2 
\eeq
(we have called $h_{+}$ and $h_{\times}$, as usual, the two independent polarization modes). After transforming to spherical coordinates, using the standard definitions  $x ^1= r \sin \theta \cos \phi$, $x^2 = r \sin \theta \sin \phi$, 
$x^3 = r \cos \theta$, we easily obtain:
\bea
&&
\chi^{rr}= \sin^2 \theta \left( h_+ \cos 2 \phi +h_\times \sin 2 \phi \right);~~
\chi^{r\theta}= {\sin 2 \theta\over 2 r } \left( h_+\cos 2 \phi+h_\times \sin 2 \phi \right);~~
\chi^{r\phi}= {1\over  r }\left( -h_+\sin 2 \phi + h_\times \cos 2 \phi \right);
\nonumber \\
&&
\chi^{\theta\theta}= { \cos^2 \theta\over r^2}  \left( h_+ \cos 2 \phi +h_\times \sin 2 \phi \right);~~
\chi^{\theta\phi}= {\cos \theta \over  r^2 \sin \theta} \left( -h_+\sin 2 \phi + h_\times \cos 2 \phi \right);~~
\nonumber \\
&&
\chi^{\phi\phi}= -{1\over r^2\sin^2 \theta} \left( h_+ \cos 2 \phi +h_\times \sin 2 \phi \right).
\eea
Since $h_{+}= h_+(\eta, r \cos \theta)$, $h_{\times}= h_\times(\eta, r \cos \theta)$, all perturbation components depend on $\phi$ only through  $\cos2 \phi$ or $\sin 2 \phi$, so that their contribution averages to zero when inserted into Eq. (\ref{A3}). 
The same is true for the case of vector perturbations, with the only difference that the $\phi$ dependence of $v^i$, in spherical coordinates, is through  $\cos  \phi$ or $\sin  \phi$ (corresponding to waves of helicity one instead of helicity two as in the tensor case).  Also the vector contribution thus averages to zero when inserted into Eq. (\ref{A3}).

%%%%%%%%%%%%%%%%%%%%%%%%%%%%%%%%%%%%%%%%%%%%%%%%%%%%%%%%%%%%%%%%%%%%%%%%%%%%%%%%%%%%%%%

\renewcommand{\theequation}{B.\arabic{equation}}
\setcounter{equation}{0}

\section*{Appendix B. Computation of the $\Ccal(\Tcal^{(1,1)}_i)$ spectral coefficients of $\overline{ \lla \Ical_{1,1} \rra }$}
\label{AppB}

We give here the result for the $\Ccal(\Tcal^{(1,1)}_i)$ spectral coefficients of $\overline{ \lla \Ical_{1,1} \rra }$ computed in the CDM case. We have introduced the convenient notation $l=k \Delta \eta$, and we have enclosed in a box the leading contributions. Finally, we have defined ${\rm Sinc} (l)= \sin (l) /l$.
\bea
\Ccal(\Tcal_{1}^{(1,1)}) &=& 0 ~~~~~,~~~~~ 
\Ccal(\Tcal_{2}^{(1,1)}) = \boxed{\Xi_s \frac{f_s^2 - f_o^2}{\Delta\eta^2} \frac{l^2}{3}} ~~~~~,~~~~~
\Ccal(\Tcal_{3}^{(1,1)}) = 0, \nonumber \\
\Ccal(\Tcal_{4}^{(1,1)}) &=& \boxed{2 \Xi_s \frac{f_s^2}{\Delta\eta^2} \frac{l^2}{3}} + \frac{4 \Xi_s}{3 l^2} \left\{ 2 - 3l^2 +( l^2 -2) \cos(l) + l \sin(l) + l^3 {\rm SinInt}(l) \right\} \nonumber \\
& & + 4 \Xi_s \frac{f_s}{\Delta \eta} ( \cos l-2+{\rm Sinc}(l)+l {\rm SinInt}(l)), \nonumber \\
\Ccal(\Tcal_{5}^{(1,1)}) &=& \boxed{- 2 \Xi_s \frac{f_o^2}{\Delta\eta^2} \frac{l^2}{3}} ~~~~~,~~~~~
\Ccal(\Tcal_{6}^{(1,1)}) = 4 \Xi_s [ 1 - {\rm Sinc}(l) ] - 4 \Xi_s \frac{f_s}{\Delta\eta} \left( \cos(l) - {\rm Sinc}(l) \right), \nonumber \\
\Ccal(\Tcal_{7}^{(1,1)}) &=& - \frac{2 \Xi_s}{3} \left\{ \boxed{\frac{f_s^2}{\Delta\eta^2} l^2} - 3 \frac{f_s}{\Delta \eta}( \cos(l) - {\rm Sinc}(l) ) - \frac{f_s}{\Delta \eta} \frac{2}{\Hcal_o \Delta\eta} [ 2 \cos(l) + ( l^2 -2)  {\rm Sinc}(l) ] \right\} ,\nonumber \\
\Ccal(\Tcal_{8}^{(1,1)}) &=& \frac{2 \Xi_s}{\Hcal_s} f_s \frac{1}{3 \Hcal_o \Delta\eta^3} \left\{ \boxed{\Hcal_o \Delta\eta ~ l^2} + 2 [- 3 \Hcal_o \Delta\eta + (l^2 - 6) ] \cos(l) - 3 ( 2 + \Hcal_o \Delta\eta ) ( -2 + l^2 ) {\rm Sinc}(l) \right\} ,\nonumber \\
\Ccal(\Tcal_{9}^{(1,1)}) &=& 0 ~~,~~
\Ccal(\Tcal_{10}^{(1,1)}) = 0~,~ 
\Ccal(\Tcal_{11}^{(1,1)}) =  8 \Xi_s \left\{\frac{1}{2}- \frac{1}{3 l^2}+\left[\frac{1}{3 l^2}-\frac{1}{6}\right] \cos(l)-\frac{1}{6}{\rm Sinc}(l)-\frac{l}{6} {\rm SinInt}(l)\right\}, \nonumber \\
\Ccal(\Tcal_{12}^{(1,1)}) &=& \left[ \Xi_s^2 - \frac{1}{\Hcal_s \Delta\eta} \left( 1 - \frac{\Hcal_s'}{\Hcal_s^2} \right) \right] \left\{ 2 ~ [1 - {\rm Sinc}(l)] + 2 \frac{f_o - f_s}{\Delta\eta} [cos(l) - {\rm Sinc}(l)] \right. \nonumber \\
& & \left. + \boxed{\frac{f_o^2 + f_s^2}{\Delta\eta^2} \frac{l^2}{3}} - \frac{2 f_o f_s}{\Delta\eta^2} [ 2 \cos(l) + ( l^2 -2) {\rm Sinc}(l) ] \right\}, \nonumber \\
\Ccal(\Tcal_{13}^{(1,1)}) &=& 4 \left\{ 1 - {\rm Sinc}(l) + \frac{f_o}{\Delta\eta}\left[cos(l) - {\rm Sinc}(l)\right] \right\} ,\nonumber \\
\Ccal(\Tcal_{14}^{(1,1)}) &=& 2\Xi_s\Big[\frac{f_o-f_s}{\Delta \eta}(\cos l- {\rm Sinc}(l)) 
+  \boxed{\frac{f_o^2}{3\Delta \eta^2}l^2} - \frac{f_of_s}{\Delta \eta^2}\left(2 \cos(l) + ( l^2 -2){\rm Sinc}(l)\right) +\left(1+2\frac{f_o+f_s}{\Delta \eta}\right)\left(1-{\rm Sinc}(l)\right)\Big], \nonumber \\
\Ccal(\Tcal_{15}^{(1,1)}) &=& 4 ~ \Xi_s \frac{f_o + f_s}{\Delta \eta} \left\{ 1 - {\rm Sinc}(l) \right\} - \frac{8}{l} {\rm SinInt}(l)~,~ 
\Ccal(\Tcal_{16}^{(1,1)}) = \frac{8}{l}{\rm SinInt}(l) ~,~
\Ccal(\Tcal_{17}^{(1,1)}) = \frac{8}{l^2} \left\{ -1 + \cos(l) + l ~ {\rm SinInt}(l) \right\}, \nonumber \\
\Ccal(\Tcal_{18}^{(1,1)}) &=& - \frac{2}{\Hcal_s \Delta \eta} \left\{ 2 - \cos(l) - {\rm Sinc}(l) - l ~ {\rm SinInt}(l) + \frac{2 f_s}{\Delta \eta} \left( 1 + \boxed{\frac{l^2}{3}} - \cos(l) - l ~ {\rm SinInt}(l) \right) + \frac{2 f_o}{\Delta \eta} \left[ 1 - {\rm Sinc}(l) \right] \right\}, \nonumber \\
\Ccal(\Tcal_{19}^{(1,1)}) &=& \frac{4}{3 l^2} \left\{ 2 - 3 l^2 + ( l^2 -2 ) \cos(l) + l \sin(l) 
+ l^3 {\rm SinInt}(l) \right\} ,\nonumber \\
\Ccal(\Tcal_{20}^{(1,1)}) &=& -\frac{2}{\Hcal_s \Delta \eta} \left\{ - \cos (l) + {\rm Sinc}(l) + \boxed{\frac{f_s}{3 \Delta \eta} l^2} - \frac{f_o}{\Delta\eta} [ 2 \cos(l) + ( l^2 - 2) {\rm Sinc}(l) ] \right\}~,~ \Ccal(\Tcal_{21}^{(1,1)}) = 4 ~ [ 1 - {\rm Sinc}(l) ], \nonumber \\
 \Ccal(\Tcal_{22}^{(1,1)}) &=&-4 + \frac{4}{3 l^2} \left[ -2(2+3l^2) + (4 + l^2) \cos(l) + l \sin(l) + l (6 + l^2) {\rm SinInt}(l) \right]~,~~~ 
\Ccal(\Tcal_{23}^{(1,1)}) = 0~. \nonumber
\eea

\end{appendix}

%%%%%%%%%%%%%%%%%%%%%%%%%%%%%%

\end{document}